\documentclass[showpacs, oneside, twocolumn, prd, amsmath, amssymb, nofootinbib, superscriptaddress]{revtex4-2}
\usepackage{cases}
\usepackage{amsmath}
\usepackage{amssymb}
\usepackage{amsfonts}
\usepackage{amssymb}
\usepackage{dcolumn}
\usepackage{bm}
\usepackage{tikz}
\usepackage{bbm}
\usepackage{graphicx}
\usepackage[percent]{overpic}
\usepackage{xcolor}
\usepackage{array}
\usepackage{subfigure}
\usepackage{multirow}
\usepackage{makecell}
\usepackage{wasysym}
\usepackage{mathtools}
\usepackage{graphicx}
\usepackage{braket}
\usepackage{hyperref}
\usepackage{mathrsfs}
\hypersetup{colorlinks=true,
	breaklinks=true,
	pdfstartview=Fit,
	linkcolor=myblue,
	citecolor=myred,
	urlcolor=blue
}
\usepackage{cleveref}

\definecolor{myblue}{RGB}{0, 100, 200}
\definecolor{myred}{RGB}{214, 39, 40}
\definecolor{mybrown}{RGB}{123, 64, 26}
\definecolor{mydarkblue}{RGB}{44, 77, 118}
\definecolor{green3}{RGB}{44,160,44}
\newcommand{\bx}{{\hbox{\boldmath $x$}}}

\newcommand{\DSR}{}
\newcommand{\Abs}[1]{|\hspace*{-1pt}|#1|\hspace*{-1pt}|}
\newcommand{\rround}{)\hspace{-2pt})}
\newcommand{\lround}{(\hspace{-2pt}(}

\newcommand{\lyfu}[1]{\textcolor{orange}{#1}}

\newcommand{\blue}[1]{{#1}}
\usepackage{ulem}
\usepackage{xcolor}
\DeclareRobustCommand{\erase}{\bgroup\markoverwith{\textcolor{red}{\rule[.5ex]{2pt}{0.4pt}}}\ULon}

\begin{document}

\title{Quantum Treatment of Black Hole Superradiance}

\author{Lingyun Fu}
\email{lf556@cam.ac.uk}
\affiliation{School of Physics, Zhejiang University}
\affiliation{Department of Applied Mathematics and Theoretical Physics, University of Cambridge,\\Wilberforce Road, Cambridge, CB3 0WA, UK}

\author{Hidetoshi Omiya}
\email{omiya@tap.scphys.kyoto-u.ac.jp}
\affiliation{Department of Physics, Kyoto University, Kyoto 606-8502, Japan}

\author{Takahiro Tanaka}
\email{t.tanaka@tap.scphys.kyoto-u.ac.jp}
\affiliation{Department of Physics, Kyoto University, Kyoto 606-8502, Japan}
\affiliation{Center for Gravitational Physics and Quantum Information, Yukawa
Institute for Theoretical Physics, Kyoto University, Kyoto 606-8502, Japan}

\author{Xi Tong}
\email{xt246@cam.ac.uk}
\affiliation{Department of Applied Mathematics and Theoretical Physics, University of Cambridge,\\Wilberforce Road, Cambridge, CB3 0WA, UK}

\author{Yi Wang}
\email{phyw@ust.hk}
\affiliation{Department of Physics, The Hong Kong University of Science and Technology,\\Clear Water Bay, Kowloon, Hong Kong, P.R. China}
\affiliation{The HKUST Jockey Club Institute for Advanced Study, The Hong Kong University of Science and Technology, Clear Water Bay, Kowloon, Hong Kong, P.R. China}

\author{Hui-Yu Zhu}
\email{hzhuav@ibs.re.kr}
\affiliation{Cosmology, Gravity and Astroparticle Physics Group, Center for Theoretical Physics of the Universe, Institute for Basic Science (IBS), Daejeon, 34126, Korea}

\begin{abstract}
Rotating black holes can form dense boson clouds through superradiant instability, making Kerr black holes a powerful probe of ultralight massive bosons. Previous studies of black hole superradiance have often treated bosonic fields classically, leaving open questions about how particles are produced and how the clouds grow over time. In this work, we canonically quantize a massive scalar field around a Kerr black hole, providing a fully quantum description of black hole superradiance. We show that the evolution of the particle number in the cloud, as well as the energy and angular momentum of the scalar field, can be consistently explained within the standard framework of quantum field theory in curved spacetime. Furthermore, we prove that the growth of the cloud occurs independently of the choice of initial state. We also explore several phenomena related to a massive scalar field in a rotating black hole spacetime, including Hawking radiation, adiabatic backreaction on the black hole spin, and the direction of level transitions in the presence of self-interactions of the field. Our analysis provides a consistent quantum-mechanical perspective that includes all these phenomena.
\end{abstract}

\maketitle

\section{Introduction}

Ultralight bosonic (ULB) particles have attracted increasing attention in recent years as one of the candidates for dark matter (DM) \cite{Press:1989id,Hu:2000ke,Goodman:2000tg,Marsh:2015xka,Hui:2016ltb}. In minimal setups, they interact with ordinary matter only through gravity, making black holes (BHs) an exceptional detection channel. One remarkable gravitational phenomenon arising from the interplay between BHs and ULBs is the BH superradiance instability, a process in which a spinning BH transfers its rotational energy to surrounding bosonic fields, generating a boson cloud. This mechanism offers a unique window into new physics \citep{zel1971generation,zel1972amplification,Starobinskii:1973vzb,Damour:1976kh, Zouros:1979iw,Detweiler:1980uk,Dolan:2007mj, Brito:2015oca}. 

The boson cloud and the BH can form a bound state analogous to that of a hydrogen atom, a system commonly referred to as the gravitational atom (GA). The GA displays a rich array of fascinating phenomena, one example being the emission of monochromatic gravitational waves (GWs) \citep{Arvanitaki:2010sy,Yoshino:2013ofa,Chan:2022dkt,Banerjee:2024nga, Siemonsen:2022yyf, Collaviti:2024mvh, Omiya:2024xlz} with frequencies set by the mass of the ultralight boson. These frequencies can fall within the sensitivity bands of current and future GW detectors \citep{Brito:2017zvb,Brito:2017wnc,Isi:2018pzk,Ghosh:2018gaw,Tsukada:2018mbp,LISAConsortiumWaveformWorkingGroup:2023arg}, making the GA an effective probe of ultralight bosons.
When the GA is embedded in a binary system, the gravitational perturbation of the companion can deform the structure of the cloud, which, in turn, backreacts on the orbital dynamics of the companion. Such deviations in the binary orbits can potentially be detected through a variety of observational channels, thereby providing the first evidence of ULBs \cite{Baumann:2019ztm,Ding:2020bnl,Takahashi:2021yhy,Tong:2021whq,Tong:2022bbl,Baumann:2021fkf,Cole:2022yzw,Takahashi:2023flk,Tomaselli:2023ysb,Fan:2023jjj,Spieksma:2024voy,Zhu:2024bqs,Tomaselli:2024dbw,Ding:2025nxe,Dyson:2025dlj,Tomaselli:2025jfo,Guo:2025pea,Ding:2025hqf}.

However, most existing studies treat the bosonic field as a classical field, where the superradiant instability is described as an exponential growth in the wave amplitude over time. In this framework, the growth of the particle population in the quantum perspective remains poorly understood. The primary challenge in quantizing the system comes from the complex frequencies of the superradiant modes, which give rise to their characteristic exponential time dependence \cite{Detweiler:1980uk,Unruh:1974bw}. This makes it difficult to apply the standard procedure of field quantization based on a set of harmonic oscillators.
Although a formal quantization approach has been attempted \cite{Ford:1975tp}, the results were incomplete due to the lacking of discrete superradiant spectra which is yet unknown at the time. 
A closely related issue i.e. the quantum treatment of ergoregion instabilities has also been explored \cite{Kang:1997uw}, offering useful conceptual guidance. In this paper, we further develop the formalism of a quantum description of superradiance along this direction.

In this work, we aim to formulate the superradiant instability in a quantum setup by applying the standard canonical quantization procedure to a massive scalar field in the Kerr metric. We present all normalizable solutions to the Klein-Gordon (KG) equation, including modes with both discrete and continuous frequencies. The discrete modes have imaginary frequencies but are normalizable differently from the quasi-normal modes (QNMs). To understand how superradiance arises in the quantum framework, we introduce a pseudo-particle number corresponding to the portion of the field localized outside the BH potential barrier. Although the particle number defined in the usual way is constant, the pseudo-particle number grows exponentially with time in the quantum description as well, regardless of the choice of initial state. This growth of pseudo-particle number can therefore be interpreted as the growth of the ``boson cloud''.

This paper is organized as follows. In Sect.~\ref{Sec.KGandMode}, we begin with the massive KG equation in the Kerr background and build a complete set of mode functions that is ortho-normalized for the mode expansion of a field. Then, we perform the canonical quantization of the massive scalar field. After quantization, in Sect.~\ref{Sec. QuantSR}, we analyze the behavior of the Hamiltonian and the angular momentum and calculate the evolution of the particle number with time. Introducing a pseudo-number operator, we show that the growth of the cloud is inevitable, whatever initial state is selected. Sect.~\ref{Sec.Discussion} then discusses related topics based on our quantum formulation, including the Hawking radiation, the adiabatic evolution of the background spacetime, and level transitions with the self-interaction turned on. Sect.~\ref{Sec.Conclusion} is devoted to the conclusion. In the Appendix, we also show the equivalence between two different expressions for the retarded Green's function in the presence of discrete modes. We adopt the $(-,+,+,+)$ metric sign convention and set $G=\hbar=c=1$ throughout our paper.

\section{Massive Scalar Field around A Kerr Black Hole}
\label{Sec.KGandMode}
\subsection{Field equations and Klein-Gordon norm}
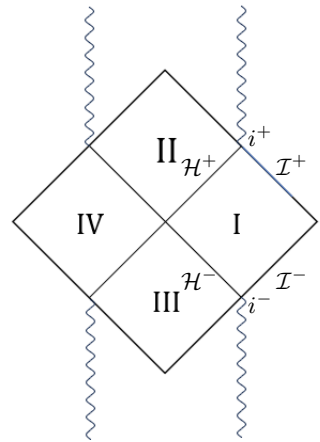
\begin{figure}[htbp]
    \centering
    \begin{overpic}[width=0.6\linewidth, percent]{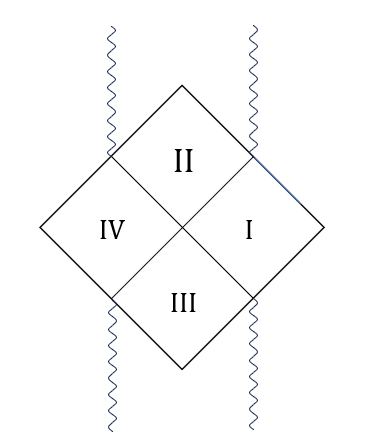}
    \put(44, 59){$\mathcal{H}^+$}
    \put(44, 35){$\mathcal{H}^-$}
    \put(64, 59){$\mathcal{I}^+$}
    \put(64, 35){$\mathcal{I}^-$}
    
    % i+ and i-
    \put(58, 65){$i^+$}
    \put(58, 30){$i^-$}
  \end{overpic}
    \caption{The Penrose diagram of Kerr BH {on the equatorial plane}. Region I is the universe we live while region IV is the parallel universe. $\mathcal{H}^\pm$ is the future/past {outer} horizon, $\mathcal{I}^\pm$ is the future/past null infinity and $i^\pm$ is the future/past timelike infinity. The wavy line corresponds to the singularity.}
    \label{fig:KerrBHPenrose}
\end{figure}

We consider a massive scalar field in a Kerr spacetime, whose metric 
is given in the Boyer-Lindquist (BL) coordinates as 
\begin{equation}
    \begin{aligned}
        {d}s^2&=-\left(1-\frac{2Mr}{\Sigma}\right){d}t^2-\frac{4Mar\sin^2\!\theta}{\Sigma}{d}t{d}\varphi+\\
        &\sin^2\!\theta\left(r^2+a^2+\frac{2Ma^2r\sin^2\!\theta}{\Sigma}\right){d}\varphi^2+\frac{\Sigma}{\Delta}{d}r^2+\Sigma {d} \theta^2,
    \end{aligned}
\end{equation}
where $\Sigma \equiv r^2 + a^2 \cos^2\!\theta$ and $\Delta \equiv r^2 - 2 M r + a^2$, with $M$ and $a$ denoting the mass and spin of the black hole, respectively. The event horizon $r_+$ and the inner horizon $r_-$ are the solutions of $\Delta = 0$, yielding $r_{\pm} = M \pm \sqrt{M^2 - a^2}$. We denote the metric tensor by 
$g_{\mu\nu}$. The Penrose diagram of the Kerr BH is shown in Fig.~(\ref{fig:KerrBHPenrose})

We consider a non-interacting real massive scalar field $\Phi$ in the Kerr spacetime, whose action is given by 
\begin{align}
{\cal S}:=\int d^4x \sqrt{-g}\,{\cal L},
\end{align}
with 
\begin{align}
{\cal L}:=\frac12 
\left[-g^{\mu\nu}(\partial_\mu \Phi)(\partial_\nu \Phi)-\mu^2\Phi^2 \right]. 
\end{align}
We expand the field operator $\Phi$ 
\if0
as 
\begin{align}
    \Phi &= \sum_{\substack{i}}(u_{i} a_{i} + u^*_{i} a_{i}^{\dagger}), 
\end{align}
by 
\fi
using a complete set of properly ortho-normalized functions with respect to the (pseudo) KG inner product defined by 
\begin{equation}
     \lround u, v \rround  \equiv i \int_{\Sigma_t}  d^3 x \sqrt{-g}\, 
     \tilde n_\mu j^\mu(u,v), 
     \label{Eq.KGInnerProduct}
\end{equation}
with {$\tilde n^\mu$ being the normal (but non-unit) vector perpendicular to the $t$-constant hypersurface, defined by $\tilde n_\mu dx^\mu = -dt$}, 
and
\begin{align}
j^\mu(u,v) := g^{\mu \nu} \left[u\, \partial_\nu v- \left(\partial_\nu  u\right) v \right], 
\label{def:j}
\end{align}
so as to satisfy 
\begin{align}
&\lround u_i, {\cal J}u_j\rround = -\lround {\cal J}u_i , u_j \rround\left\{
\begin{array}{l}
 \ne 0,\quad (\mbox{for}~~ i= j),\cr
 = 0,\quad (\mbox{for}~~ i\ne j),
\end{array}\right.\cr
&\lround u_i, u_j\rround = \lround {\cal J}u_i , {\cal J}u_j \rround =0.
\label{orthnormalization of modes}
\end{align}

Usually, the conjugation operation ${\cal J}$ corresponds to complex conjugation, but in our analysis, we adopt a different definition {which simultaneously flips the sign of $\omega$ and $m$ in the eigenmodes,} in order to apply it to the discrete superradiant modes. {It turns out that such a sign-flipping definition of $\cal J$ is {equivalent to} the complex conjugation for the usual continuous modes} \cite{Cannizzaro:2023jle} . 
In the usual KG inner product,  $v$ on the right-hand side in Eq.~\eqref{Eq.KGInnerProduct} is replaced with its complex conjugate, $u_j^*$, but we adopt this unconventional notation because the ordinary convention turns out to be not so convenient when we discuss discrete superradiant modes. 
{Because} the KG inner product is independent of the choice of the hypersurface,  here we give an expression which is valid only when the hypersurface is set to be a constant-time hypersurface, for simplicity. 

The equation of motion for the massive scalar field in Kerr spacetime is given by  \cite{Teukolsky:1973ha}
\begin{equation}
    \Box \Phi = \mu^2 \Phi ~ , 
    \label{Klein-Gordon eq}
\end{equation}
where the d'Alembertian operator $\Box = \nabla^\nu \nabla_\nu$ and $\mu$ is the mass of the scalar field. 
As the Kerr metric is independent of $t$ and $\varphi$, one can assume the separation of variables with respect to $t$ and $\varphi$. We can further separate $r$ and $\theta$, owing to the symmetry of the Kerr space-time~\cite{Brill:1972xj}, to obtain a complete set of modes in the form of 
\if0
\begin{equation}
    u_{\omega j m} = e^{- i \omega t} e^{i m \varphi} 
    U_{\omega j m}(r,\theta), 
    \label{ansatz}
\end{equation}
where $j$ is an additional label to discriminate between different 
solutions with the same values of $\omega$ and $m$. 
Assuming the separation of variables $U_{\omega j m}(r,\theta)
= R(r) S(\theta)$, 
\fi
\begin{equation}
    u_{\omega l m} = e^{- i \omega t} e^{i m \varphi} R_{\omega l m}(r) S_{\omega l m}(\theta)~.
    \label{ansatz}
\end{equation}
The radial part and the polar part of the equation are separated as
\begin{align}
   {\cal L}_\theta S_{\omega l m}(\theta)&:=\biggl[\frac{{d}}{{d} \cos\theta }\sin ^2\theta \frac{{d}}{{d} \cos\theta}
    - \frac{m^2}{\sin^2 \theta} + 
    2am\omega
   \cr 
    &\!\!\!\!\!\!
     - a^2(\omega^2\sin^2\theta + \mu^2 \cos^2 \theta) + \lambda_{lm\omega}\biggr] S_{\omega l m}(\theta) = 0 ~ , 
\label{polar equation}\\
   {\cal L}_r R_{\omega l m}(r)
    &:=\biggl[\frac{{d}}{{d} r}\Delta \frac{{d}}{{d} r} 
    + \frac{\left(\omega (r^2 + a^2) - a m\right)^2}{\Delta} \cr
    & \qquad\qquad -\mu^2 r^2  - \lambda_{lm\omega}\biggr]R_{\omega l m}(r)= 0 ~ , 
\label{radial equation}
\end{align}
where $\lambda_{lm\omega}$ is a separation constant, which is determined by the regularity of the spheroidal harmonics $S(\theta)$ at both 
$\theta=0$ and $\pi$. 
Here, we introduce an integer label $l \ge |m|$ such that $S_{lm\omega}$ reduces to the spherical harmonics $Y_{lm}$ in the non-rotating limit $a \to 0$. 
We normalize the spheroidal harmonics to satisfy the orthogonality conditions 
\begin{equation}
    \int_0^\pi{d}\theta\, \sin\theta \, S_{\omega lm}(\theta) S_{\omega l'm}(\theta)  = \frac{1}{2\pi}\delta_{ll'} ~ , 
\end{equation}
and these conditions specify the normalization of $S_{\omega lm}(\theta)$. 
We should note that there is a symmetry
\begin{align}
    S_{\omega lm}(\theta)=
S_{-\omega l\,-m}(\theta),
\label{Ssym}
\end{align} 
because Eq.~\eqref{polar equation} is invariant 
under the exchange $(\omega,m)\to (-\omega\,-m)$. 

The radial function generally takes the asymptotic forms 
\begin{align}
       R_{\omega l m}(r) \rightarrow {\cal N}_{\omega l m} \mathcal G(r)
       \left\{
       \begin{array}{l}
	 A_{\omega l m}^{\rm out} e^{ik_Hr_*}
     +A_{\omega l m}^{\rm in}e^{- ik_Hr_*},  \cr
      \qquad\qquad\qquad\qquad (r_*\to-\infty),\cr
	 A_{\omega l m}^{\rm up}e^{ik_\infty r_*}
     + A_{\omega l m}^{\rm down}e^{-ik_\infty r_*},  \cr
      \qquad\qquad\qquad\qquad  (r_*\to+\infty),
	\end{array}\right. 
    \label{AsymBehavior}
\end{align}
Here, the tortoise coordinate $r_*$ is defined by ${d}r_* := (r^2 + a^2) dr/\Delta$. The horizon $r= r_+$ and infinity $r= +\infty$ corresponds to $r_* = -\infty$ and $r_* = +\infty$, respectively.  
The coefficients $A^{\rm in}, A^{\rm out}, A^{\rm up},$ and $A^{\rm down}$ are, in general, complex numbers. We denote a normalization factor as ${\cal N}$, which we demand to be real and positive for a real $\omega$. Determination of ${\cal N}$ will be done later.
In the asymptotic form~\eqref{AsymBehavior}, we introduced
\begin{align}
 &   \mathcal G(r):= 1/\sqrt{r^2+a^2}~,\\ 
 &  k_\infty(\omega):=\sqrt{\omega^2-\mu^2}~,\\
 &   k_H(\omega,m):=\omega-m\Omega_H~,\\
 & \Omega_H = \frac{a}{2Mr_+}~.
\end{align} 
In this paper, we take the branch cuts of $k_\infty$ to emanate from $\omega=\pm \mu$ with an infinitesimal negative imaginary part, extending towards infinity in the lower half complex plane.

In the present setup, the inner product 
$\lround  u_{\omega lm}, u_{\omega' l'm'}  \rround$
is non-vanishing only when $\omega'=-\omega$ and $m'=-m$. 
If the inner product with $\omega'+\omega\ne 0$ is non-vanishing, 
it should depend on time in proportion to $\exp[-i(\omega+\omega')t]$, 
which contradicts the fact that the KG inner product is time-independent.
%constant. 
By inserting the Kerr metric, the inner product is 
expressed explicitly as
\begin{align}
   \lround  u_{\omega lm}, u_{\omega' l'm'}  \rround & =  
   \delta_{l,l'}\delta_{m,-m'}\cr
 &   \times \int {d}r_* \,  \frac{f(r)}{{\cal G}^2} 
    R_{\omega lm}(r)R_{\omega' lm}(r),
\label{Eq.Normu}
\end{align}
where we introduce the weight of the radial integral 
\begin{align}
        f(r) \equiv 2\omega\left[1- \frac{\eta\, a^2 \Delta}{(r^2 + a^2)^2}\right] - \frac{4 a M r}{(r^2 + a^2)^2} m ~ ,
        \label{Eq:fofr}
\end{align}
with 
\begin{align}
    \eta \equiv 2 \pi\int d(\cos\theta) \sin^2\!\theta\, [S_{\omega lm}(\theta)]^2. 
\end{align}
Here, we used the fact that the inner product vanishes unless $\omega=-\omega'$.
The weight function behaves asymptotically as 
\begin{align}
f(r)\to \left\{
\begin{array}{ll}
2\omega, & (r_*\to\infty),\cr
2 k_H, & (r_*\to -\infty).
\end{array}
   \right. 
   \label{eq:Asymptotic_weight}
\end{align}

\subsection{Complete set of mode functions}

In this section, we list all the normalizable modes with either continuous or discrete frequency spectra, in preparation for the mode expansion discussed later.

\subsubsection{Continuous modes}

\begin{figure}[htbp]
        \centering
        \begin{overpic}[width=0.9\linewidth, percent]{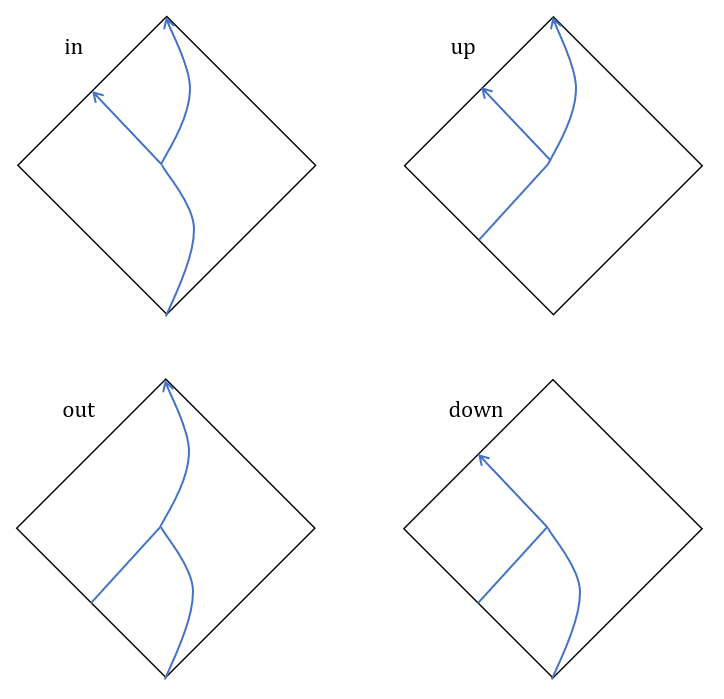}
      \end{overpic}
        \caption{All four kinds of modes in 
        {the region I of Kerr BH Penrose diagram shown in Fig.~\ref{fig:KerrBHPenrose}}. The ``in'' and ``up'' modes represent the wave ingoing and upcoming from $i^-$ and $\mathcal{H}^-$ respectively, while the ``out'' and ``down'' modes represent the wave outgoing and down to $i^+$ and $\mathcal{H}^+$ respectively.}
        \label{fig:allmodes}
    \end{figure}

\blue{We start with the usual mode solutions with $\omega_I=0, |\omega_R|>\mu$, where $\omega_I$  and $\omega_R$ are 
the imaginary and real parts of $\omega$, respectively. 
In this case, for each set of values of $(\omega, \ell, m)$, we have two independent radial solutions, which form two independent continuous spectra. 
In principle, discrete modes with $\omega_I = 0$ could appear, but for the system considered here, no such modes exist.}  
It would be convenient to choose ``in'' mode 
$R^{\rm in}_{\omega l m}$ and ``up'' mode $R^{\rm up}_{\omega l m}$, 
which are specified by the asymptotic boundary conditions
\begin{equation}
    \begin{aligned}
    R^{\rm in}_{\omega l m}(r) \to {\cal N}_{\omega l m}^{\rm in}
            \mathcal G\begin{dcases}
     e^{- ik_Hr_*}, & (r_*\to -\infty)\\
     A^{\rm up}_{\omega l m}e^{+ ik_\infty r_*} &
     \\
    \qquad +A^{\rm down}_{\omega l m}e^{-ik_\infty r_*},  & (r_*\to+\infty),
    \end{dcases}
    \label{Eq.RadialEqin}
    \end{aligned}
\end{equation}
and
\begin{equation}
    \begin{aligned}
         R^{\rm up}_{\omega l m}(r) \to
         {\cal N}_{\omega l m}^{\rm up}\mathcal G\begin{dcases}
    A^{\rm in}_{\omega l m } e^{-ik_Hr_*} 
    \\
    \qquad
    +A^{\rm out}_{\omega l m }e^{+ik_Hr_*},  & (r_*\to -\infty),\\
    e^{+ ik_\infty r_*}, & (r_*\to+\infty).
    \end{dcases}
    \end{aligned}
    \label{DefRup}
\end{equation}
The ``in'' mode represents a scalar wave ingoing from $i^-$, which is partly reflected to $i^+$ while partly transmitted to $\mathcal{H}^+$. The ``up'' mode represents a scalar wave upcoming from $\mathcal{H}^-$, which is partly reflected to $\mathcal{H}^+$ while partly transmitted to $i^+$. The ``in'' and ``up'' modes are depicted in Penrose diagram shown in Fig.~\ref{fig:allmodes}.
We denote the corresponding apparently ``positive'' frequency functions by 
$u^{\rm in}_{\omega l m}$ and $u^{\rm up}_{\omega l m}$, respectively.  

For $\omega_I=0$ and $0<|\omega_R|<\mu$, $k_\infty$ becomes pure imaginary and 
${\rm Im} \, k_\infty >0$. We chose the branch cut of $k_\infty$ such that 
$R^{\rm up}_{\omega l m}$ satisfies the decaying boundary condition at 
$r_*\to +\infty$. 
In this frequency range we only have 
one normalizable mode $R^{\rm up}_{\omega l m}$,
since 
$R^{\rm in}_{\omega l m}$ contains the exponentially growing component at $r_*\to +\infty$. 

In the region $|\omega_R| >\mu$ we can define ``out'' and ``down'' modes that satisfy the boundary conditions shown in Fig.~\ref{fig:allmodes}, 
by flipping the sign of 
$\omega$ and $m$ in $R^{\rm in}_{\omega l m}(r)$
and $R^{\rm up}_{\omega l m}(r)$. 
Under this sign-flipping operation, which we denote by ${\cal J}$, the radial equation \eqref{radial equation} remains invariant. 
In addition, the radial equation \eqref{radial equation} is real when $\omega_I=0$, and 
the solutions are real for $|\omega_R| < \mu$ except for the possibility of multiplying an irrelevant overall phase.

We assume that the normalization factors ${\cal N}_{\omega l m}$ are all real by appropriately choosing the irrelevant overall phases. 
Since 
${\cal J}k_H=-k_H$ and ${\cal J}k_\infty=-k_\infty$ hold 
in the region $|\omega_R|>\mu$, we find {that we can consistently impose the condition}
\begin{align}
 {\cal J}^{\Re}   R^{\rm x} =&
  (R^{\rm x})^*, ~~~\mbox{for}~|\omega|>\mu,~{\rm x=``in",``up"}, 
  \label{Eq:JRx}
\end{align}
where we associate 
the superscript {$\Re$} to ${\cal J}$ to express that $\omega$ is above the branch cut in the complex plane of $\omega$. 
If we naively flip the sign of $\omega$ with $\omega_I>0$, it should be mapped to the value below the complex plane in general. 
We continue to use ${\cal J}$ to express this transformation. 

On the real axis satisfying $|\omega|<\mu$, threre is no distinction between ${\cal J}$ and ${\cal J}^{\Re}  $. 
For $|\omega_R|< \mu$, 
${\cal J}k_\infty=k_\infty$ and the radial function is real, and therefore we find 
\begin{align}
 {\cal J} R^{\rm up} =
   R^{\rm up} 
   = (R^{\rm up})^*,\quad \mbox{for}~|\omega|<\mu,  
\end{align}
where the first equality holds for an arbitrary $\omega$. 
Hereafter, we abbreviate the subscript $\omega lm$ unless it is ambiguous. 
To be consistent with the above transformation laws, we should choose the normalization constants as 
\begin{align}
{\cal J}^{\Re}   {\cal N}^{\rm x}=
{\cal N}^{\rm x}. 
\label{N relation}
\end{align}

Thus, the mode functions {are} defined by 
\begin{align}
 u^{\rm x}:= e^{-i\omega t+im\varphi} R^{\rm x}(r)S(\theta) ~ .
 \label{Defu}
\end{align}
{Because $S(\theta)$ is always real for the continuous modes, combining with Eq.~\eqref{Ssym},} we find 
\begin{align}
 {\cal J}^{\Re}   u^{\rm x} =
  (u^{\rm x})^*,~~(\mbox{for all continuous modes}). 
  \label{eq.Juustar}
\end{align}
Hence, we can use the the operation ${\cal J}^{\Re}  $, 
in place of the complex conjugation. 
Use of ${\cal J}^{\Re}  $ is advantageous in the sense 
that we can avoid that $\omega^*$ enters in the 
expression that requires analytic continuation 
away from the real axis \blue{to ensure that the expressions are holomorphic when analytically continued away from the real axis}. 

From the above mentioned relation, we can read 
the relations among the asymptotic coefficients:
\begin{align}
& {\cal J}^{\Re}   A^{\rm x} =
 (A^{\rm x})^*,~~~(\mbox{for}~|\omega|>\mu),
 \label{AstaronBC}\\
& {\cal J}A^{\rm out} =
 A^{\rm in}=(A^{\rm out})^*, \quad
 (\mbox{for}~ |\omega|<\mu). 
 \label{JArelation}
\end{align}
The first equality in \eqref{JArelation} generically holds for an arbitrary $\omega$. 
We introduce $\Abs{A^{\rm x}}^2:=A^{\rm x}
({\cal J}^{\Re} A^{\rm x})$.
From Eqs.~\eqref{AstaronBC} and \eqref{JArelation}, we find $\Abs{A^{\bf x}}^2 = |A^{\bf x}|^2$ for all continuous modes.

When both $R_1(r)$ and $R_2(r)$ are solutions of 
the radial equation, the Wronskian 
\begin{align}
W[R_1(r),R_2(r)]\equiv \frac1{{\cal G}^{2}} \left(
   R_1(r) \frac{dR_2(r)}{dr_*} 
   - \frac{dR_1(r)}{dr_*} R_2(r) 
   \right),
\end{align}
becomes a constant independent of $r_*$. 
Then, from the conservation of
$W[R^{\rm in},{\cal J}R^{\rm in}], W[R^{\rm up},{\cal J}R^{\rm up}],$ and $W[R^{\rm up},{\cal J}R^{\rm in}]$ at the infinity and the horizon,
we have the identities,
\begin{align}
& k_\infty =k_H \left(\Abs{A^{\rm out}}^2 -
\Abs{A^{\rm in}}^2\right),
\label{k1}\\
& k_H =k_\infty \left(\Abs{A^{\rm down}}^2 -
\Abs{A^{\rm up}}^2\right), 
\label{k2}\\
& k_\infty {\cal J}^{\Re}   A^{\rm up}
 =  -k_H A^{\rm in} ,  
\label{k4}
\end{align}
for $|\omega|>\mu$, and  from $W[R^{\rm up},R^{\rm in}],$ we have
\begin{align}
& k_\infty A^{\rm down}
 =k_H A^{\rm out},
\label{k3}
\end{align}
for an arbitrary $\omega$. 

Now, we are ready to evaluate the inner products 
between $u^{\rm x}$ and ${\cal J} u^{\rm x}$. 
After straightforward computation, we find~\cite{Balakumar:2022yvx,Frolov:1998wf}, 
\begin{align}
& \lround u^{\rm up}_{\omega lm}, {\cal J}^{\Re}   u^{\rm up}_{\omega' lm}\rround\cr
& \quad =2\pi (\mathcal N^{\rm up})^2 
    \left(k_\infty+k_H 
     \left(\Abs{A^{\rm out}}^2+\Abs{A^{\rm in}}^2\right)
    \right)\delta(\omega-\omega')\cr 
    &\quad 
= 4\pi {(\mathcal N^{\rm up})^2 } k_H \Abs{A^{\rm out}}^2 \delta(\omega-\omega'),\cr
& \lround u^{\rm in}_{\omega lm}, 
{\cal J}^{\Re}   u^{\rm in}_{\omega' lm}\rround\cr
& \quad 
  =2\pi (\mathcal N^{\rm in})^2 
    \left(k_\infty
    \left(\Abs{A^{\rm up}}^2+\Abs{A^{\rm down}}^2\right)+k_H
    \right) \delta(\omega-\omega')
    \cr&\quad 
= 4\pi {(\mathcal N^{\rm in})^2} k_\infty \Abs{A^{\rm down}}^2  \delta(\omega-\omega'),
\end{align}
and the inner products of all the other pairs among 
$u^{\rm in}$, $u^{\rm up}$, ${\cal J}^{\Re}   u^{\rm in}$ and ${\cal J}^{\Re}   u^{\rm up}$ vanish. 
We can also see that 
$\lround u^{\rm in}, {\cal J}^{\Re}   u^{\rm up}\rround=0$ 
more intuitively by sending the surface to evaluate the inner product to 
$\mathcal{I}^- \cup i^- \cup \mathcal{H}^-$. 
Then, the ``in'' modes vanish at $\mathcal{H}^-$ and $\mathcal{I}^-$ while the ``up'' modes vanish at  $i^-$ and ${\cal I}^-$. Hence, they do not have any overlap on the surface $\mathcal{I}^- \cup i^- \cup \mathcal{H}^-$. 

By choosing the normalization constants as
\begin{align}
& {\cal N}^{\rm up}=\frac{1}{2\sqrt{\pi \Abs{A^{\rm out}}^2}}~, \cr
& {\cal N}^{\rm in}=\frac{1}{2\sqrt{\pi \Abs{A^{\rm down}}^2}}~, 
\label{Nnorm_conti}
\end{align}
which satisfy Eq.~\eqref{N relation},  the ortho-normalization relations among the 
continuous modes are summarized as 
\begin{align}
\lround u^{\rm up}_{\omega l m}, 
        {\cal J}^{\Re}   u^{\rm up}_{\omega' l' m'} \rround  &=  {k_H}(\omega,m)\,\delta(\omega - \omega')\delta_{ll'}\delta_{mm'}  ~ ,\cr
        \lround u^{\rm in}_{\omega l m}, 
        {\cal J}^{\Re}   u^{\rm in}_{\omega' l' m'} \rround &= k_\infty(\omega)\,
        \delta(\omega - \omega')\delta_{ll'}\delta_{mm'},
\label{OrthoNormContinuous} 
\end{align}
with all other inner products being 0 \cite{Balakumar:2022yvx,Frolov:1998wf}. 
\blue{Whether the mode is of positive or negative frequency is determined by the signature of the norm $k_H(\omega , m)$ and $k_\infty(\omega)$.} 
Hence, from Eq.~\eqref{OrthoNormContinuous}, we find that when $k_H > 0$, the mode $u^{\rm up}$ is the positive frequency mode, while $\mathcal{J}^{\Re} u^{\rm up}$ corresponds to the negative frequency mode, as implied by the relation in Eq.~\eqref{orthnormalization of modes}. Conversely, when $k_H < 0$, the roles are reversed, suggesting $u^{\rm up}$ becomes the negative frequency mode, while $\mathcal{J}^{\Re} u^{\rm up}$ becomes the positive frequency mode.

In the same manner, $u^{\rm in}$ is the positive frequency function for ${k_\infty} >0$, {\it i.e.},
for $\omega >\mu$, while ${\cal J}^{\Re}   u^{\rm in}$ is the positive frequency function in the opposite case. 
We did not include $1/{\sqrt{|k_H|}}$ and $1/{\sqrt{|k_\infty|}}$ in the normalization 
coefficients ${\cal N}$, to keep all the expressions 
holomorphic. 
We associate the factor $k_H$ and $k_\infty$ 
with the commutation relation between the creation and annihilation operators later.  
To summarize, only for the superradiant modes, $u^{\rm up}$ with $0<\omega<m\Omega_H$, the roles of positive and negative frequency modes are interchanged between 
$u^{\rm up}$ and ${\cal J}^{\Re}   u^{\rm up}$ than the ordinary case. 
We denote the class of continuous modes in this range by $N^-$, 
while the class of the other continuous modes by $N^+$, {\it i.e.},
\begin{align}
N^-\!: & ~\{u^{\rm up}_{\omega lm},{\cal J}^{\Re}   u^{\rm up}_{\omega lm}\}~\mbox{ with } m>0~\mbox{and}~0<\omega<m \Omega_H,\hspace{-10mm}\cr
N^+\!: & ~ \mbox{all the other continuous modes}. 
\end{align}
 It is worth pointing out that observers floating near the horizon, 
 whose 4-velocity $u^\mu$ should satisfy 
 $u^\mu\partial_\mu\propto 
 \partial_t + \Omega_{H}\partial_\varphi $ approximately, would identify 
 ${\cal J}^{\Re}   u^{\rm up }_{\omega l m}$ with $k_H<0$ as a positive frequency mode without any doubt, because
\begin{align}
 iu^\mu\partial_\mu \log({\cal J}^{\Re}   u^{\rm up }_{\omega l m})\vert_{\mathcal{H}^-}= -k_H >0.  
\end{align}
\blue{As $u^{\rm up}$ and ${\cal J}^{\Re}   u^{\rm up}$ vanish at $i^-$ and ${\cal I}^-$, by considering 
the limit $t\to -\infty$, this argument explains why ${\cal J}^{\Re}    u^{\rm up}_{\omega lm}$ with $k_H(\omega,m)<0$ is the positive frequency function.}

\subsubsection{Superradiant discrete modes}
To construct a complete set of mode functions, we also need to consider the discrete spectrum, 
which satisfy the decaying boundary conditions on both boundaries at $r_*=\pm\infty$. 
We find that these boundary conditions imply 
$A^{\rm out}=0$ and $A^{\rm down}=0$ simultaneously at discrete frequencies with $\omega_I>0$, which corresponds to the modes that undergo the superradiant instability. Their amplitude grows exponentially with time by extracting rotational energy from the BH \citep{Damour:1976kh, Zouros:1979iw,Detweiler:1980uk,Dolan:2007mj, Brito:2015oca}.

Normalizable solutions with discrete complex frequencies with $\omega_I>0$ {can be obtained through} the analytic continuation of 
$u^{\rm up}$ (or $u^{\rm in}$) given in Eq.~\eqref{Defu} with Eq.~\eqref{Eq.RadialEqin} (or Eq.~\eqref{DefRup}). 
In the case of the analytic continuation of $u^{\rm up}$, the boundary condition at the horizon 
is automatically satisfied 
as long as $\omega_I>0$, and the decaying boundary 
condition toward the horizon requires that 
the frequencies of discrete modes $\omega=\bar\omega_{nlm}$ must satisfy  
\begin{align}
  A^{\rm out}_{\bar\omega_{nlm} lm}=0. 
\end{align}
We assign $n=l+1$ to the mode that possesses the smallest $\omega_R$, in analogy with Hydrogen atom. 
We can discuss the analytic continuation of $u^{\rm in}$ in a similar manner, and we obtain the equivalent condition for the frequency, $A^{\rm down}_{\bar\omega_{nlm} lm}=0$, {\it c.f.}, Eq.~\eqref{k3}. 

Thus, the radial function labeled by $nlm$
is expressed as 
\begin{align}
    R_{nlm}(r) \to	 {\cal N}^{\rm DSR}_{\bar\omega_{nlm} lm}
       {\mathcal G}\begin{dcases}
        A_{\bar\omega_{nlm} l m}^{\rm in} e^{- ik_Hr_*},  & (r_* \to -\infty),\\
        e^{+ ik_\infty r_*},  & (r_* \to +\infty). 
        	\end{dcases} 
            \label{Eq.RQNM}
\end{align}
These modes are quite different from the 
quasi-normal modes (QNMs). 
The QNMs are also specified by the zeros of $A^{\rm out}$. However, the QNMs are decaying, {\it i.e.}, $\omega_I<0$.
For the modes not to vanish completely at a large $|r_*|$ at a late time, they exponentially grow radially towards the both boundaries at $r_*\to \pm \infty$. 
Therefore, $\omega$ is not on the Riemann surface that we are currently concerned with, on which 
${\rm Im} k_\infty \geq 0$ always holds, but on another one beyond the branch cuts. Anyway, because of their exponential growth at $r_*\to \pm \infty$, QNMs are not normalizable. 
{Nevertheless, they appear as poles in the retarded Green's function, corresponding to 
$A^{\rm out}=0$. We give an explicit form of the retarded Green's function written as a single contour integral with respect to $\omega$ in Appendix~\ref{Sec:retardedGF}, where we show that the retarded Green's function in the form of a single contour integral is equivalent to the one composed of the mode sum of all normalizable modes. The QNM poles are in the Riemann surface beyond the branch cut, instead of in the one that we mainly consider in this paper. When we focus on the late-time behavior, we can close the integration contour in the lower-half complex $\omega$-plane, and those QNM poles are picked up, as usual. }
We do not discuss QNMs in this paper any further, because they have nothing to do with the mode expansion of the field operator, $\phi$.
For more details about QNMs, see \cite{Berti:2009kk}. 

The superradiant discrete mode functions are obtained as 
\begin{align}
  u^{\DSR}_{nlm}=e^{-i\bar\omega_{nlm}t+im\varphi}
     R_{nlm}(r)S_{\bar\omega_{nlm}lm}(\theta). 
\end{align}
Since the KG equation Eq.~\eqref{Klein-Gordon eq} is real and symmetric under the sign-flipping operation ${\cal J}$, we also have three more independent solutions, 
\begin{align}
    u^{\DSR *}_{nlm}, 
{\cal J}u^{\DSR}_{nlm} 
\mbox{ and } {\cal J}u^{\DSR *}_{nlm}, 
\label{eq40}
\end{align}
\blue{where the action of ${\cal J}$ is the flip of the signs of $\bar\omega_{nlm}$ and $m$.} In particular, the last one, 
\begin{align}
{\cal J}u^{\DSR *}_{nlm} =e^{-i\bar\omega^*_{nlm}t+im\varphi}
     R^*_{nlm}(r)S^*_{\bar\omega_{nlm}lm}(\theta), 
\end{align}
belongs to the same $m$-modes as the original mode $u^{\DSR}$, and its frequency is given by $\bar\omega^*$. 
Thus, we find that two modes, $u_{nlm}$ with $\omega=\bar\omega_{nlm}$ and $\mathcal{J}u^*_{nlm}$ with $\omega=\bar\omega_{nlm}^*$, always appear in a pair. 
The frequency of the former is in the upper half complex plane while that of the latter is in the lower-half complex plane. 
We should recall Eq.~\eqref{Ssym} and that we can extend the relation on the real axis, 
\begin{align}
 {\cal J} R^{\rm up}_{\omega lm}=R^{\rm up}_{\omega lm},  
 ~~~\mbox{for} ~~\omega_I= 0~\mbox{and}~ |\omega_R|<\mu, 
 \label{Eq:JR}
\end{align}
to the whole complex plane consistently, because 
\begin{align}
 k_\infty(-\omega)=k_\infty(\omega), 
  ~~~\mbox{for} ~~\omega_I\ne 0~\mbox{or}~ |\omega_R|<\mu.   
\end{align}
To be consistent with Eq.~\eqref{Eq:JR}, 
we should have 
\begin{align}
    {\cal N}_{\bar\omega lm}^{\rm DSR}={\cal J}{\cal N}_{\bar\omega lm}^{\rm DSR}. 
    \label{eq:Nsymmetry}
\end{align}
We should stress that ${\cal J}R_{nlm}(r)$ does not take the form of Eq.~\eqref{Eq.RQNM}. 
The asymptotic form of ${\cal J}R_{nlm}(r)$ at $r_*\to -\infty$ is 
$\propto A^{\rm out}_{-\bar\omega_{nlm}l-m}e^{ik_H(-\bar\omega_{nlm},-m)}$ and $A^{\rm in}_{-\bar\omega_{nlm}l-m}=0$. 

We consider the KG inner product among these discrete modes. Since the KG inner product must be time-independent, from the vanishing conditions of $t$- and $\varphi$-dependence, we find that the non-vanishing combination is essentially $\lround u_{nlm}, {\cal J} u_{nlm} \rround$ only. 
As given in the Appendix \ref{Appendix:DiscreteNorm}, one can calculate 
the inner product as
\blue{
\begin{align}
 \lround u_{nlm},{\cal J} u_{nlm} \rround
 =2i ({\cal N}_{\bar\omega lm}^{\rm DSR})^2
  k_H A^{\rm in}_{\bar\omega lm} \partial_{\bar\omega}A^{\rm out}_{\bar\omega lm}. 
\end{align}
Here, we abbreviate the subscripts $nlm$ from $\bar\omega_{nlm}$, for brevity. 
Hence, we set the normalization coefficient to 
\begin{align}
 {\cal N}^{\rm DSR}_{n lm} 
 =\left[2i k_H 
 A^{\rm in}_{\bar\omega lm} \partial_{\bar\omega}A^{\rm out}_{\bar\omega lm}
 \right]^{-1/2},
 \label{calNDSR}
\end{align}
which is consistent with Eq.~\eqref{eq:Nsymmetry}.  
With this normalization, we obtain
\begin{align}
 \lround u_{nlm},{\cal J} u_{n'l'm'} \rround
&=\lround {\cal J}u^*_{nlm},u^*_{n'l'm'} \rround\cr
& 
 =\delta_{nn'}\delta_{ll'} \delta_{mm'},
 \label{discrete_norm}
\end{align}
while all other inner products of discrete modes vanish.}

We should stress that when we express the complex conjugate $u^*_{nlm}$, we can eliminate all complex conjugate operations except for the one acting on $\bar\omega_{nlm}$. 
Namely, 
\begin{align}
 & \hat R^{\rm up*}_{\bar\omega lm}
= \hat R^{\rm up}_{\bar\omega^* lm}, \cr
& {A^{\rm in *}_{\bar\omega lm} = A^{\rm out}_{\bar\omega^* lm},\quad A^{\rm in}_{\bar\omega^* lm} = A^{\rm out *}_{\bar\omega lm} = 0},\cr
&    R^*_{nlm}(r)={\cal N}^{\rm DSR*}_{nlm}
        \hat R^{\rm up*}_{\bar\omega lm},\cr
&      {\cal N}^{{\rm DSR}*}_{n lm} 
 =\left[-2i k_H(\bar\omega^*,m) 
 A^{\rm out}_{\bar\omega^* lm} \partial_{\bar\omega^*}
  A^{\rm in}_{\bar\omega^* lm}\right]^{-1/2},\cr
& S^*_{\bar\omega lm}(\theta)
 =  S_{\bar\omega^* lm}(\theta),
\end{align}
{where the first equality is derived from the fact that $k_\infty^{*}(\bar \omega) = -k_\infty(\bar \omega^*)$, and all other equalities can be derived directly.

It would be worth mentioning how the orthogonality 
\begin{align}
\lround u^{\DSR}_{nlm}, u^{\DSR *}_{nlm}\rround=0,
\label{uustar}
\end{align}
is satisfied. 
This orthogonality relies on the orthogonality of the radial functions:
\begin{align}
\int dr_* \tilde f(r) (r^2 + a^2) |R_{nlm}|^2=0,
\label{Eq:OrthogonalityDSR}
\end{align}
with
\begin{align}
    \tilde f(r) := 2\omega_R\left[1- \frac{\eta\, a^2 \Delta}{(r^2 + a^2)^2}\right] - \frac{4 a M r}{(r^2 + a^2)^2} m.
\end{align}
Eq.~\eqref{Eq:OrthogonalityDSR} is explained by the sign change in the weight function $\tilde f(r)$, which is related to the presence of the ergoregion. 
The outer region of the radial integral in Eq.~\eqref{Eq:OrthogonalityDSR} contributes positively, while the inner region contributes negatively. 

One example of the radial function $R_{211}(r)$ is shown in Fig.~\ref{fig:Rofu}.
\begin{figure}[htbp]
    \centering
    \includegraphics[width=0.8\linewidth]{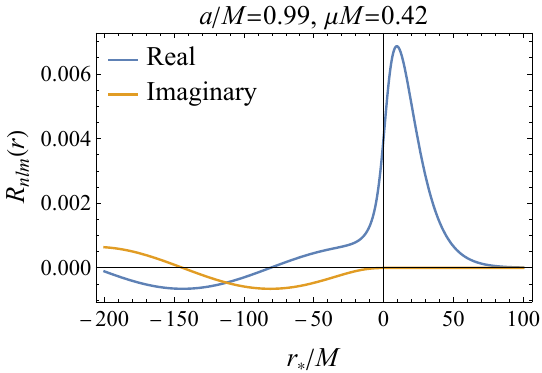}
    \caption{{Radial profile $R_{211}(r)$ of the fundamental mode. The BH and field parameters are set as $a/M = 0.99, \mu M = 0.42$. The blue line denotes the real part of $ R_{211}$ while the orange line denotes the imaginary part.}}
    \label{fig:Rofu}
\end{figure}
The bump in the real part of the radial function is on the outside of the potential barrier, which is the part that we usually identify as the axion cloud. 
From this plot, we find that the cancellation in the integral in Eq. \eqref{Eq:OrthogonalityDSR} is caused by a long tail of $|R_{nlm}|^2$ inside the barrier, although its amplitude is suppressed.
As a result, the inner product of $\lround u_{nlm},u^{*}_{nlm}\rround$ vanishes.
The integral $\int dr_* (r^2+a^2)f(r) R_{nlm}(r)^2$, which is contained in $\lround u_{nlm}, {\cal J} u_{nlm}\rround$, is dominated by the region outside the potential barrier, because $R_{nlm}(r)^2$ oscillates inside the barrier \blue{and gives only a negligible small contribution.}

\subsection{Mode expansion}
As we have obtained a complete set of normalizable solutions with the evaluation of the KG norm in the previous subsection, we expand the field as 
\blue{
\begin{align}
\Phi = \sum_{l}&\Biggl[\sum_{m=1}^l
  \sum_{\substack{n = l+1}}^{\infty}\Biggl(
 u_{n lm} a_{n lm} + u_{n lm}^* a_{n lm}^{\dagger} \cr
    &\qquad + \mathcal{J}u_{nlm}^* b_{n l m} 
    + \mathcal{J}u_{nlm} b^{\dagger}_{n l m}\Biggr) ~ \cr
    &+\sum_m \Biggl(\int_{-\infty}^\infty
    u^{\rm up}_{\omega lm } a^{\rm up}_{\omega lm} {d}\omega\cr
     &\quad + \left(
     \int_{-\infty}^{-\mu}+\int^{\infty}_{\mu} \right)
    u^{\rm in}_{\omega lm } a^{\rm in}_{\omega lm}{d}\omega\Biggr)\Biggr].
    \label{Eq:mode_decomposition1}
\end{align}}
Now, we establish the commutation relation among the expansion coefficients introduced above. 
Since $\Phi$ is Hermitian, together with Eq.~\eqref{eq.Juustar} we have 
\begin{align}
a^{{\rm x}\dag}_{\omega lm}={\cal J}^{\Re} a^{\rm x}_{\omega lm}=a^{\rm x}_{-\omega\, l\,-m}. 
\end{align}
We can extract the expansion coefficients as 
\begin{align}
  &a_{nlm}=\lround \Phi, {\cal J}u_{nlm}\rround, 
  \cr
  &
  a^{\rm up}_{\omega lm}=
  k_H^{-1}\lround \Phi, {\cal J}^{\Re} u^{\rm up}_{\omega lm}\rround, \cr
  & a^{\rm in}_{\omega lm}=k_\infty^{-1}\lround \Phi, {\cal J}^{\Re} u^{\rm in}_{\omega lm}\rround, 
\end{align}
Using the equal-time commutation relations
 \begin{align}
& [\Phi(t,\boldsymbol{x}), \Pi(t,\boldsymbol{x'})] = i \delta^{(3)}(\boldsymbol{x} - \boldsymbol{x'}) ~ ,  \cr
& [\Pi(t,\boldsymbol{x}), \Pi(t,\boldsymbol{x'})] = 0 ~ , \cr
& [\Phi(t,\boldsymbol{x}), \Phi(t,\boldsymbol{x'})] = 0 ~ ,
 \end{align}
we obtain the desired commutation relations as
\begin{align}
        [a_{nlm}, b^\dagger_{n' l' m'}] &= 
        [b_{nlm}, a^\dagger_{n' l' m'}]
        =\delta_{nn'}\delta_{ll'}\delta_{mm'}  ~ ,
        \label{nonstarndardCR}\\ 
        [a_{\omega lm}^{\rm up}, a^{\rm up\dag}_{\omega' l' m'}] &=  \frac{1}{k_H(\omega,m)} \delta(\omega - \omega')\delta_{ll'}\delta_{mm'}  ~ , \\
        [a_{\omega lm}^{\rm in}, a^{\rm in\dag}_{\omega' l' m'}] &=  \frac{1}{k_\infty(\omega)} \delta(\omega - \omega')\delta_{ll'}\delta_{mm'}  ~ , 
\end{align}
and the commutators of all the other combinations vanish. We introduced the conjugate momentum $\Pi = \partial\mathcal{L}/\partial\dot{\Phi}$.
Here, we used $a^{{\rm x}\dag}_{\omega l m}={\cal J}^{\Re} a^{\rm x}_{\omega l m}$ for continuous modes. 
As anticipated before, the commutation relation of the associated expansion coefficients has an unusual frequency-dependent factor.  
Because of this factor, $\sqrt{|k_H|}a_{\omega lm}^{\rm up}$ and $\sqrt{|k_\infty|}a_{\omega lm}^{\rm in}$ are the annihilation operators normalized in a standard notation for $k_H>0$ and for $k_\infty >0$, respectively. They plays the role of the creation operator in the opposite case. We stick to the current notation in this paper, because the expression becomes more compact and the use of $|k_H|$ hides the analyticity of expressions. 

The commutation relations for the discrete modes differ from the standard case. To obtain the standard one, we introduce new operators $a_{\pm,nlm}$ by   
\begin{equation}
    \left\{
    \begin{aligned}
        a_{nlm} &= \frac{1}{\sqrt{2}} (a_{+,nlm} + a_{-,nlm}^\dagger),  \\
        b_{nlm} &= \frac{1}{\sqrt{2}} (a_{+,nlm}-a_{-,nlm}^\dagger), ~ 
    \end{aligned}
    \right.
    \label{Eq.apm}
\end{equation}
such that satisfy the ordinary commutation relations: 
\begin{equation}
\begin{aligned}
    [a_{+,n l m}, a^\dagger_{+,n' l' m'}] &= \delta_{nn'} \delta_{ll'} \delta_{mm'} ~ , \\
    [a_{-,n l m}, a^\dagger_{-,n' l' m'}] &= \delta_{nn'}  \delta_{ll'} \delta_{mm'} ~ . 
\end{aligned}
\label{4.4.6}
\end{equation}
The mode functions associated with the new operators $a_{\pm,nlm}$ are given by
\begin{align}
v_{+,nlm} &= \frac{1}{\sqrt 2}(u_{nlm} + \mathcal {J} u^*_{nlm}),\cr
v_{-,nlm} &= \frac{1}{\sqrt 2}(u^*_{nlm} - \mathcal {J} u_{nlm}). 
\end{align}
It is easy to check that they obey the following ortho-normalization relations: 
\begin{align}
  &  \lround v_{\pm, nlm}, {\cal J}v_{\pm, nlm}\rround =  1, \cr
  &  \lround v_{\pm, nlm}, {\cal J}v_{\mp, nlm}\rround = \lround v_{\pm, nlm}, {\cal J}v^*_{\pm, nlm}\rround = 0.
\end{align}
The mode expansion is now reorganized as
\begin{align}
    \Phi =& \sum_{n l}\sum_{m=1}^l
     \bigl(v_{-,n lm} a_{-,n lm} + v_{-,n lm}^* a_{-,n lm}^{\dagger} \nonumber\\[-3mm]
    &\qquad\qquad +v_{+, nlm} a_{+,n l m}
    + v^*_{+,nlm} a^\dagger_{+,n l m} \bigr) ~ \cr
    &+ (\mbox{continuum contributions with } \omega_I=0). 
    \label{Eq.ModeExpansionNew}
\end{align}

Since ${\cal J}(R_{nlm}S_{nlm})=R_{nlm}S_{nlm}$,
we find 
\begin{align}
  v_{\pm,nlm}\approx \frac{e^{\mp i\omega_R t\pm im\varphi}}{\sqrt{2}}
  \left({R^*_{nlm}}{S^*_{nlm}}\pm R_{nlm}S_{nlm}\right),
\end{align}
assuming $\omega_I t\ll 1$. Since $\omega_I$ is small, 
$S_{nlm}$ is almost real. 
Hence, the real (imaginary) part of $R_{nlm}$ corresponds to $v_+$ ($\lyfu{-}v^*_-$) at $t=0$.  
As is read from Fig.~\ref{fig:Rofu}, $v_+$ has contribution also inside the 
potential barrier, but the contribution from 
the outside dominates, while $v_-$ is concentrated in the region inside the barrier.

\section{Quantum dynamics of Superradiance}
\label{Sec. QuantSR}
In this section, we discuss quantum states and their properties. 
In the present setup we do not have a vacuum state as the lowest energy state, which makes the procedure unconventional. 

\subsection{Hamiltonian and the reference state}
\label{Sec.Hamiltonian}

The Hamiltonian is given by
\begin{align}
    H &\equiv \int_{\Sigma_t} d^3x \sqrt{-g} (\Pi \dot{\Phi}-{\cal L}) \cr
    &=\int_{\Sigma_t} d^3 \boldsymbol{x} \sqrt{-g}
    \Bigl(-\frac{1}{2} g^{00} \partial_0 \Phi \partial_0 \Phi \cr
    &\qquad \qquad+ \frac{1}{2} g^{ij} \partial_i \Phi \partial_j \Phi 
    + \frac{1}{2}\mu^2 \Phi^2\Bigr).
    \label{HamiOriginal}
\end{align}
We can show that this Hamiltonian is equivalent to 

\begin{align}
    E &:= 
    \int_{\Sigma_t} d^3 x \sqrt{-g} T_{\mu\nu} \tilde n^\mu \xi^\nu,
        \label{Hami1}
\end{align}
where 
\begin{align}
 T_{\mu\nu}\equiv 
 {\partial_{(\mu} \Phi \partial_{\nu)} \Phi} -\frac12 g_{\mu\nu}
 \left(\partial_\rho\Phi \partial^\rho\Phi +\mu^2\Phi^2\right),
\end{align}
and $\xi^\mu$ is the time-like Killing vector defined by $\xi^\mu\partial_\mu=\partial_t$.
We notice that $\tilde n_\mu \xi^\mu=-1$. 
Parentheses on the indices mean the symmetrization. 
\blue{
Performing the integration by parts in the spatial directions with Dirichlet boundary conditions at finite values of $r_*$, simplifying the expression using the KG equation, and sending the boundary to infinity ($r_*\to \pm\infty$) later, we obtain
\begin{align}
  H= &\int _{\Sigma_t} d^3 x \sqrt{-g}
   \Bigl(\xi^\nu {\tilde n}^\mu  \partial_{(\mu} \Phi
    \partial_{\nu)} \Phi \cr
   &- \frac{1}{2} {\tilde n}_\nu\xi^\nu 
   g^{00} \left(\Phi \partial^2_0 \Phi + \partial_0 \Phi \partial_0 \Phi\right)\Bigr). 
        \label{Hami2}
\end{align}}
Plugging the mode expansion shown in Eq.~\eqref{Eq:mode_decomposition1}, 
the second term in Eq.~\eqref{Hami2} and the Hamiltonian reduces to
\begin{align}
        H = H_{\rm D} + H_{\rm C},
\end{align}
with
\begin{align}
H_{\rm D}        
&:= \sum_{n lm} \bar\omega_{nlm} 
  b_{nlm}^\dagger a_{nlm} 
  +\bar\omega_{nlm}^* 
  a_{nlm}^\dagger  b_{nlm},\\ 
H_{\rm C}& :=\sum_{lm} \Biggl[
   \int_\mu^\infty 
    d\omega\, \omega k_\infty ({\cal J}^{\Re}  a^{\rm in}_{\omega l m})a^{\rm in}_{\omega l m} \cr
    &\qquad \qquad
    + \int_0^\infty d\omega\, \omega k_H
    ({\cal J}^{\Re}  a^{\rm up}_{\omega l m})a^{\rm up}_{\omega l m} \Biggr]~ , 
\label{Eq.HamiOperatorDC}
\end{align}
besides an additive constant irrelevant to the current discussion.  $H_{\rm D}$ represents the contribution of discrete modes, and $H_{\rm C}$ is that of continuous modes. 
This is nothing but $\sum_i \omega_i n_i$ with $n_i$ being the number operator of the $i$-th mode. 

The expression for the angular momentum along the direction of the axis of Kerr metric, $J_z$, is obtained by replacing $\xi^\mu$ in Eq.~\eqref{Hami2} with the minus of the rotational Killing vector $-\xi^\mu_{(\varphi)}$ defined by 
{$\xi^\mu_{(\varphi)}\partial_\mu=\partial_\varphi$}, and hence the expression of $J_z$ in the mode decomposition is obtained by simply replacing $\omega$ in \eqref{Eq.HamiOperatorDC} with $m$, {\it i.e.},
\begin{align}
        J_z = J_{z\rm D} + J_{z\rm C},
\end{align}
with
\begin{align}
J_{z\rm D}        
&:= \sum_{n lm} m\left( 
  b_{nlm}^\dagger a_{nlm} 
  +  a_{nlm}^\dagger  b_{nlm}\right),\\ 
J_{z\rm C}& :=\sum_{lm} m \Biggl[
   \int_\mu^\infty 
    d\omega\, k_\infty ({\cal J}^{\Re}  a^{\rm in}_{\omega l m})a^{\rm in}_{\omega l m} \cr
    &\qquad \qquad
    + \int_0^\infty d\omega\, k_H
    ({\cal J}^{\Re}  a^{\rm up}_{\omega l m})a^{\rm up}_{\omega l m} \Biggr]~ . 
\label{Eq.JOperatorDC}
\end{align}
}

The Hamiltonian of continuous modes, $H_C$, is represented as two independent series of modes, {\it i.e.}, ``in'' and ``up''. 
``Up'' modes contain oscillators belonging to $N^-$, which have a negative energy. Hence, $H_C$ is unbounded from below. 
Recall that we have to identify $a^{\rm up}_{\omega l m}$ not as the annihilation operator but as the creation operator for the modes in $N^-$. 
This interchange of the roles of the creation and annihilation operators is necessary to ensure the positivity of the norm of the excited states. 
Therefore, it is more appropriate to replace the Hamiltonian shown in Eq.~(\ref{Eq.HamiOperatorDC}) for the modes in $N^{-}$, excluding their zero-point energy, by 
\begin{align}
 H_C(N^-)=\sum_{l=1}^\infty \sum_{m=1}^l \int_0^{m\Omega_H} d\omega\, (-\omega)|k_H|
    a^{\rm up}_{\omega l m} ({\cal J}^{\Re}  a^{\rm up}_{\omega l m}) , 
\end{align}
though this is just an unessential shift of the origin of the energy. 

Because the Hamiltonian is unbounded from below, we do not have the Boulware ``vacuum", the lowest-energy state.  
However, one can still define a Boulware-like reference state $|B\rangle_C$ by
the conditions 
\begin{align}
  a_{\omega lm}\ket{B}_C=0,&\qquad \mbox{for modes in}~N^+,\cr
  ({\cal J}^{\Re}  a_{\omega lm})\ket{B}_C = 0,&\qquad \mbox{for modes in}~N^-, 
\end{align} 
and construct the Fock space $\mathcal{H}_C$ applying creation operators to it. 
\blue{When there is no interaction between two sectors $N^+$ and $N^-$, they can be treated as two independent quantum systems. Then, the flip of the sign of the energy is just a matter of convention and the reference state $\ket{B}_C$ is stable.}
However, $\ket{B}_C$ becomes unstable once the interaction between the two sectors is activated.
Nonetheless, as long as free field theory is concerned, there is no pathological behavior associated with this negativity of the excitation energy, and $\ket{B}_C$ is stable. 
The quantum state $\ket{B}_C$ is also singular on the horizon, as in the case of the usual Boulware vacuum state in a non-rotating BH spacetime. 
We will discuss later about the states that are regular on the future horizon

Next, we discuss the Hamiltonian for the discrete modes, $H_D$. 
Together with Eq.~\eqref{Eq.apm} and Eq.~\eqref{4.4.6}, $H_{\rm D}$ can be rewritten as 
\begin{align}
    H_D \equiv H_0 + H_S,
\end{align}
with 
\begin{align}
    H_0 & := \sum_{n lm} \bar\omega_{R}(a_{+} ^\dagger a_{+} - a_{-}^\dagger a_{-}) ~ ,\cr
    H_S & := \sum_{n lm} i\bar\omega_{I}(a_{+}^\dagger a_{-}^\dagger - a_{+}a_{-}) ~ . 
\label{Eq:HamiltonianDiscrete}
\end{align}
Since the imaginary part of the superradiant discrete modes frequency is small, {\it i.e.}, $\bar\omega_I\ll \bar\omega_R$, $H_0$ is the dominant part while $H_S$ is a small correction to it. 
Then, the modes $v_{+,nlm}$ have a positive energy, while the modes $v_{-,nlm}$ have a negative energy. 
Because of the negative energy modes, $H_D$ is also unbounded from below, and hence the lowest energy state does not exist. Note that this type of Hamiltonian appears as a non-degenerate parametric amplifier in the quantum optics~\cite{Scully:1997hcl}. 

Unlike the case of continuous modes we discussed above, because of the presence of $H_S$, the positive- and the negative-energy sectors interact with each other. 
\blue{Therefore, it is not only that the lowest-energy state does not exist, but that {normalizable} energy eigenstates themselves are absent.}
It is easy to prove that $H_D$ cannot be rewritten into the diagonalized form, in which the Hamiltonian takes the form of $H_D=\sum \omega_i a^\dag_i a_i$. 
If it were the case, we should have $[a_i,H_D]=\omega_i a_i$. 
However, as $a_i$ should be a linear combination of $a$, $b$, $a^\dag$ and $b^\dag$, {\it i.e.}, $a_i=\alpha a+\beta b+\gamma a^\dag+ \delta b^\dag$, a direct computation shows that $[a_i,H_D]=-\alpha \bar\omega^* a -\beta \bar\omega b+\gamma \bar\omega^* a^\dag+\delta \bar\omega b^\dag$, which is incompatible with the condition $[a_i,H_D]=\omega_i a_i$ mentioned above. 
This completes the proof of the absence of the basis in which the positive and negative energy sectors decouple.

We specify a reference state for the discrete modes $\ket{B}_D$ by the conditions 
\begin{align}
  a_{+,nlm}\ket{B}_D = a_{-,nlm}\ket{B}_D = 0,
\end{align}
and then we construct the Fock space $\mathcal{H}_D$ multiplying the creation operators to it. 
Such a reference state is not invariant under the time translation, since the annihilation operators $a_{\pm}$, which defines the reference state, do not commute with the Hamiltonian, {\it i.e.}, the operator for the time translation. 

One may think that we can specify the reference state by imposing the conditions $a_{nlm}\ket{B}_D = b_{nlm}\ket{B}_D = 0$, but these conditions are equivalent to $a_{+}\ket{B}_D = a^\dag_{-}\ket{B}_D = 0$. 
Then, the excited states generated by multiplying $a_{-}$ (and $a^\dag_{+}$) can have a negative norm. 
Therefore, we cannot accept this choice of the construction of the Hilbert space. 
The complete Hilbert space should be $\mathcal{H}_C \otimes \mathcal{H}_D$, and we choose the joint reference state as $\ket{B} \equiv \ket{B}_C \otimes \ket{B}_D$.

{Since the number operators associated with discrete modes $v_{\pm,nlm}$,
\begin{align}
    N_{\pm,nlm} &= a_{\pm,nlm} ^\dagger a_{\pm,nlm},
    \label{Eq:NumOp}
\end{align}
do not commute with the Hamiltonian, the corresponding Heisenberg operators $N_{\pm}(t)$ evolve in time obeying the Heisenberg equation, $dN_{\pm}(t)/dt=-i[H,N_{\pm}(t)]$. The evolutions of $N_{\pm}(t)$ are easily solved to obtain 
\begin{align}
N_+(t)+&N_-(t)=2\sinh^2(\bar\omega_I t) \cr
    &-\frac{\sinh(2\bar\omega_I t)}{2} \left(a_{+}^\dag(0) a_{-}^\dag(0) + a_{+}(0) a_{-}(0)\right)\hspace{-5mm}\cr
 &+\cosh(2\bar\omega_I t)(N_+(0) +N_-(0)),
    \label{Eq:Nppm} \\
 N_+(t)-&N_-(t)= N_+(0) - N_-(0) ~ .\label{Eq:Npmm}
\end{align}
The exponentially growing factors in Eq.~\eqref{Eq:Nppm} roughly explains the growth of superradiant clouds, which we will discuss in more detail in the next subsection, while Eq.~\eqref{Eq:Npmm} reflects the conservation of energy.}

\subsection{Growth of superradiant clouds}
\label{sec:growth_of_superrad}
The expectation value of the Hamiltonian $H$ is time-independent for any quantum state, since $H$ is the generator of time translation and it commutes with itself. This is nothing but the energy conservation, which is the consequence of the time translation symmetry of the background Kerr spacetime. 
\blue{We should notice that each time slicing must be a Cauchy surface in quantum field theory. Otherwise, we cannot describe the unitary evolution of the quantum state. Hence, the escape of the energy through the boundary is strictly prohibited.} 
On the other hand, in the situation where the superradiant condensate grows, the energy of the scalar field also looks growing. 
How can we understand the growth of the clouds under the constraint of the energy conservation?

In order to understand the growth of superradiant condensates in the language of the quantum field theory, we first need to realize that the amplitude of the mode functions $v_\pm$, roughly speaking, depends on time as $|v_{\pm}|\propto \cosh(\omega_I t)$. 
This is because $v_\pm$ are composed of a superposition of exponentially growing modes, $u_{nlm}, u^*_{nlm} \propto e^{\omega_I t}$ and exponentially decaying modes, {${\cal J}u_{nlm},{\cal J}u^*_{nlm}\propto e^{-\omega_I t}$. }

The second step is to understand how the KG norm of these mode functions remains constant and the energy is conserved as well. 

We explain how this cancellation occurs. 
For this purpose, we divide the contribution of the inner product into two pieces as 
$\lround *,*\rround =\lround *,*\rround_{\rm int}+\lround *,*\rround_{\rm ext}$, where the inner product with the subscript ``int'' (``ext'') represents the contribution of the region inside (outside) the potential barrier. 
Because of the oscillatory behavior of the 
product of the radial mode functions $R^2_{nlm}$ \blue{in the internal region}, we have 
\begin{align}
 &\lround u^{\DSR}_{nlm},
  {\cal J} u^{\DSR}_{nlm}\rround_{\rm int}\approx 0,
  \label{intvanishment}
\end{align}
and hence we have 
\begin{align}
\lround u^{\DSR}_{nlm},
  {\cal J} u^{\DSR}_{nlm}\rround_{\rm ext}  \approx 1, 
\end{align}
which shows the radial mode function $R_{nlm}$is almost real in the external region as is read from Fig.~\ref{fig:Rofu} explicitly.
This means that ${\cal N}^{\rm DSR}_{nlm}$ is also almost real positive number. 
Using this fact, we have 
\begin{align}
  e^{-2\bar\omega_I t}\lround u^{\DSR}_{nlm},
   u^{\DSR *}_{nlm}\rround_{\rm ext}
& \approx \lround u^{\DSR}_{nlm},
  {\cal J} u^{\DSR}_{nlm}\rround_{\rm ext}  \approx 1. 
  \label{ID2}
\end{align} 
On the other hand, Eq.~\eqref{uustar} implies
\begin{align}
 \lround u^{\DSR}_{nlm},u^{\DSR *}_{nlm}\rround_{\rm int}
 =- \lround u^{\DSR}_{nlm},u^{\DSR *}_{nlm}\rround_{\rm ext}. 
 \label{IPrelations}
\end{align}

Since $\bar\omega_I$ is small, we also have 
\begin{align}
& \lround u^{\DSR}_{nlm},
  {\cal J} u^{\DSR}_{nlm}\rround_{\rm int/ext}
  \approx 
   \lround {\cal J}u^{\DSR *}_{nlm},
  u^{\DSR *}_{nlm}\rround_{\rm int/ext},\cr
&  
 e^{-2\bar\omega_I t}\lround u^{\DSR}_{nlm},
   u^{\DSR *}_{nlm}\rround_{\rm int/ext}
  \approx 
e^{2\bar\omega_I t}   \lround {\cal J}u^{\DSR *}_{nlm},
  {\cal J} u^{\DSR}_{nlm}\rround_{\rm int/ext}. \cr
\label{complexconjugateID}
\end{align}
Using these approximate relations \eqref{complexconjugateID}, we find 
\begin{align}
 \lround v_{\pm},{\cal J} v_{\pm}
  &   \rround_{\rm int/ext}
 \approx
 \lround u^{\DSR}_{nlm},
  {\cal J} u^{\DSR}_{nlm}\rround_{\rm int/ext}\cr
  &\pm\frac{1+e^{-4\bar\omega_I t}}{2} \lround u^{\DSR}_{nlm},
   u^{\DSR *}_{nlm}\rround_{\rm int/ext}.
\end{align}
Applying the relations \eqref{intvanishment}-\eqref{IPrelations}, we obtain 
\begin{align}
\lround v_{\pm},{\cal J} v_{\pm}
    \rround_{\rm int}
 \approx &\, 0 \mp \cosh(2\bar\omega_I t),\cr 
\lround v_{\pm},{\cal J} v_{\pm}
    \rround_{\rm ext}
 \approx &\, 1 \pm \cosh(2\bar\omega_I t). 
\end{align}
Consequently, we recover 
\begin{align}
\lround v_{\pm}, {\cal J}v_{\pm}\rround = 1.
\end{align}
Since the ergoregion is in the internal region, \blue{it is consistent that} the energy of $v_-$ modes, which are more localized in the internal region, is negative. The result is consistent with the discussion at the end of Sec.~\ref{Sec.Hamiltonian}.

To discuss the growth of the condensate, it would be better to introduce the number operator corresponding to each axion cloud quasi-eigenstate. 
As there is a potential barrier in the radial equation for superradiant modes, we can find an approximate solution that stay localized in the external region for a short time.  
To describe these quasi-localized modes, we introduce a new mode function, which satisfies the initial conditions at $t=t_0$:
\begin{align}
&    \tilde u^{(t_0)}_{nlm}(t_0,\bx):=
    u_{nlm}(0,\bx)\,\theta(r-r_b),\cr
&    \dot{\tilde u}{}^{(t_0)}_{nlm}(t_0,\bx):=
    \dot u_{nlm}(0,\bx)\,\theta(r-r_b), 
    \label{temporalSRModes}
\end{align}
where $r_b$ is the point under the potential barrier at which the internal and external regions are separated. Since we mostly perform approximations in the following discussions, an accurate definition for $r_b$ is unnecessary.
We associate the annihilation and creation operators, $\tilde a^{(t_0)}_{nlm}$ and $\tilde a^{(t_0)\dag}_{nlm}$ with the mode functions $\tilde u^{(t_0)}_{nlm}(t_0,\bx)$ and $\tilde u^{(t_0)*}_{nlm}(t_0,\bx)$, respectively.
These mode functions are, of course, solutions of the KG equation, but their meaning as the modes for the superradiant condensate is limited to $t\approx t_0$. 
We define the {pseudo-}number operator 
\begin{align}
    \tilde N^{(t_0)}_{nlm}:=\tilde a_{nlm}^{(t_0)\dag}\tilde a_{nlm}^{(t_0)}, 
\end{align}
which counts the number of particles
contained in the condensate labeled by $nlm$ at $t=t_0$ \blue{in the usual sense}. 

Then, it is straightforward to calculate
\begin{align}
 \tilde a_{nlm}^{(t_0)} =& \lround \Phi(t,\bx), \tilde u^{(t_0)*}_{nlm}\rround
 =
 \lround \Phi(t+t_0,\bx), u^{*}_{nlm}\rround_{\rm ext}\cr
 = &\, 
 e^{-i\bar\omega t_0} a_{nlm} \lround u_{nlm}, u^*_{nlm}\rround_{\rm ext}\cr
 &\,  \qquad + e^{-i\bar\omega^* t_0} b_{nlm} \lround {\cal J}u^*_{nlm}, u^*_{nlm}\rround_{\rm ext}\cr
 &\,  +
e^{i\bar\omega^* t_0} a^\dag_{nlm} \lround u^*_{nlm}, u^*_{nlm}\rround_{\rm ext}\cr
 &\,  \qquad + e^{i\bar\omega t_0} b^\dag_{nlm} \lround {\cal J}u_{nlm}, u^*_{nlm}\rround_{\rm ext}\cr
\approx &\, 
 e^{-i\bar\omega t_0} a_{nlm} 
    +e^{-i\bar\omega^* t_0} b_{nlm}\cr
= &\,  \sqrt{2} e^{-i\bar\omega_R t_0}
   \left( a_+\cosh(\bar\omega_I t_0) 
        +a_-^\dag \sinh(\bar\omega_I t_0) 
        \right),\cr
\end{align}
where $\lround\cdots \rround_{\rm ext}$ is assumed to be evaluated at $t=0$ as before. 
\blue{Here, we neglect the contributions from the continuous modes in the third equality, because their contribution to the expectation value of $\bra{B}\tilde N_{nlm}^{(t_0)}\ket{B}$ is $t_0$-independent.} 
Then, the expectation value of $\tilde N^{(t_0)}_{nlm}$ is easily evaluated as 
\begin{align}
\bra{B}\tilde N_{nlm}^{(t_0)}\ket{B}
= 2\sinh^2(\bar\omega_I t_0).
 \label{eq:exp_particleNum}
\end{align}

From the expression \eqref{eq:exp_particleNum}, we find that the particle number $\bra{B}\tilde N_{nlm}\ket{B}$ increases for the reference state as the absolute value of time $|t|$ increases. We also find that $\bra{B}\tilde N_{nlm}^{(0)}\ket{B} =0$ for the reference state, \blue{besides the possible time-independent contribution from the continuous modes}. Since 
the choice of the origin of time is arbitrary, we find that there 
exists a quantum state such that $\langle\tilde N_{nlm}^{(t_0)}\rangle =0$ for \blue{each value of an arbitrarily chosen $t_0$}. 

To prove that the increase of the expectation value of the number of particles contained in a cloud, it would be enough to observe that, for an arbitrary state $\ket{\psi}$, we have
\begin{align}
    \bra{\psi}\tilde N_{nlm}^{(\delta t)}&-2\tilde N_{nlm}^{(0)}+\tilde N_{nlm}^{(-\delta t)}\ket{\psi}\cr
 &   =
4\sinh^2(\bar\omega_I \delta t)
 \bra{\psi}a^\dag_+ a_+
 + a^\dag_- a_-+1\ket{\psi}\cr
& \geq 
 4\sinh^2(\bar\omega_I \delta t),   
\end{align}
\blue{where we neglect the additional contribution from the continuous modes, again. The above inequality implies
\begin{align}
  &\biggl\langle\psi\biggl\vert
    \frac{d^2\tilde N_{nlm}^{(t_0)}}{dt_0^2}\biggr\vert \psi\biggr\rangle {\bigg|_{t_0 = 0}} \cr
   &\qquad = \lim_{\delta t\to 0}\frac{\bra{\psi}\tilde N_{nlm}^{(\delta t)}-2\tilde N_{nlm}^{(0)}+\tilde N_{nlm}^{(-\delta t)}\ket{\psi}}{\delta t^2}\cr
   &\qquad \geq {4}\bar\omega^2_I, 
\end{align}}
{which can be generalized to any $t_0$.}
Hence, whatever initial state one may choose, the growth of the superradiant condensates is inevitable. 

Finally, we briefly discuss how the discrete superradiant modes extract energy and angular momentum from the central black hole.  
For this purpose, we evaluate the radial component of the energy flux at a finite value of $r_*$, which is given by 
\begin{align} 
&  T_{\mu\nu}{\hat r}^\mu \xi^\nu =
  \partial_{(\mu}\Phi\partial_{\nu)}\Phi
 \hat r^\mu \xi^\nu,
\end{align}
where {$\hat r_\mu$ is the unit normal $ \propto \partial r/\partial x^\mu$.} 
Taking the expectation value in the region near the horizon, we can evaluate the energy flux passing through $r=$constant surface  in a unit coordinate time interval of $t$ \blue{in the increasing direction of $r$} as
\blue{
\begin{align}
\bra{B}& T_{\mu\nu} \ket{B} \hat r^\mu \xi^\nu
 \cr
 & \to -\frac12 \sum_{nl}\sum_{m>1}^l 
    (\bar\omega k_H^*+
      \bar\omega^* k_H) \left(|u_{nlm}|^2-|{\cal J}u_{nlm}|^2\right)\cr
 & =\sum_{nl}\sum_{m>1}^l 
  \left(m\bar\omega_R \Omega_H-|\bar\omega|^2 \right)
  \left({\cal N}_{\bar\omega lm}^{\rm DSR}\right)^2 |A^{in}_{\bar\omega lm}|^2 \cr
  & \qquad \ \times
   \left(e^{2\,{\rm Im}\, \bar\omega_{nlm} (t+r_*)} 
      -e^{2\,{\rm Im}\, \bar\omega_{nlm} (-t+r_*)}\right).
\label{eq:positiveFlux}
\end{align}}
The energy flux near $H^+$ is basically positive,  as long as 
\begin{align}
    m\bar\omega_R \Omega_H-|\bar\omega|^2 >0,
\label{Cond:SRcomp}
\end{align}
which is the extension of the super-radiance condition to a complex value of $\omega$~\cite{Dolan:2007mj}. 

\blue{
In the near horizon limit the wave vector $k_\mu$, which is determined by the normal vector of the phase of the horizon in-going wave, is given by 
\begin{align}
 k_\mu dx^\mu \to -\omega dt - k_H dr + m d\varphi.  
\end{align}
Raising the subscript of $k_\mu$ as
\begin{align}
  k^\mu \partial_\mu \to 
  \frac{(2Mr_+)^2 k_H}{\Sigma\Delta} 
    \left[\partial_t-\partial_{r_*}
    +\Omega_H\partial_\varphi \right], 
\end{align}
we can confirm that the wave vector is almost parallel to the ingoing null vector, independently of the value of $\omega$. Hence, the propagation direction of the superradiant discrete modes is also in-going for all modes. 
The energy flux in \eqref{eq:positiveFlux},
pointing toward the increasing direction of $r$, is the consequence of the flow of the negative energy toward the horizon.
}

\section{Discussions} 
\label{Sec.Discussion}
\subsection{Hawking radiation}
\label{Sec:HawkingRad}

So far, we consider the Boulware vacuum state as the reference state, but it is known to be singular at the horizon and physically irrelevant~\cite{Candelas:1980zt}. 
As a substitute, the Hartle-Hawking vacuum, which represents the thermal equilibrium state, is often 
considered. 
However, a more physically relevant state will be the Unruh state~\cite{Unruh:1976db}, in which the Hartle–Hawking vacuum~\cite{Hartle:1976tp} is assigned only to the “up’’ modes, which are non-vanishing at ${H}^-$. 
The ``in'' modes are kept unchanged to respect the natural vacuum choice at ${\cal I}^-$ for massless case, or at $i^-$ for the massive case. 
We can do the same in the present setup for the ``up'' modes, but the situation for the discrete modes is different. 
In Sec.~\ref{sec:growth_of_superrad} we have observed the unbounded growth of the quasi-particle number regardless of the choice of quantum state, which proves that the discrete modes do not have any stationary state. 
Namely, there is no thermal equilibrium state, {\it i.e.}, no Hartle-Hawking-like state, for discrete modes.
Therefore, it does not make much sense to search for the natural state analogous to the usual Unruh vacuum for discrete modes. 
Furthermore, since the discrete modes decay exponentially at large distances, they do not contribute to the radiation observed by a distant observer at any finite time. Consequently, the standard calculation of Hawking radiation can be extended to massive fields without any essential modification.

To identify the initial quantum state, including the discrete modes, after the formation of a rotating black hole via a gravitational collapse, it would be necessary to evolve the initial vacuum state in the dynamical spacetime, which is beyond the scope of this paper. 
However, we can imagine that the state at a time $t=t_0$ immediately after the formation of a rotating BH will not have a very large occupation number of quasi-particles, $\tilde N^{(t_0)}_{nlm}$, and the state of the continuous modes will be approximated by the Unruh state.  

Since we have found that the discrete modes do not contribute to Hawking radiation, we focus on the continuous modes in the following discussion. 
The singular behavior at the horizon is caused by the ``up" modes, but the treatment of these modes is well understood, see~\cite{Unruh:1974bw,Frolov:1998wf}. Here, we would recapitulate the non-trivial point different from the non-rotating case~\cite{Hawking:1975vcx, Ottewill:2000qh}. 
When considering only observables outside the horizon, the Unruh state is obtained by replacing the quantum state of the ``up'' modes to the thermal state with respect to $|k_H|$, which is the energy observed in the very vicinity of the horizon. 

The Hawking radiation is purely described by the ``up" modes, and the number spectrum observed by the observers at a large distance is simply given by
the product of the grey-body factor $1/\Abs{A^{\rm out}}^2$ and the expectation value of the number of the outgoing particles near the horizon.
Hence, we have 
\begin{align}
  I_\omega:=
  \frac1{\Abs{A^{\rm out}}^2}
  \bra{H} a^{\rm up\dag}_{\omega lm} 
      a^{\rm up}_{\omega lm}\ket{H} =\frac{\Abs{A^{\rm out}}^{-2}}
    {k_H(e^{\beta k_H}-1)}~,
    \label{Hawkingspectrum}
\end{align}
where the state $\ket{H}$ is the Hartle-Hawking state.
The number spectrum vanishes for $|\omega| < \mu$, as $u^{\rm up}(r)$ decays at a large $r$ exponentially in this case. Namely, the above formula applies only for $|\omega| > \mu$. 

The derivation of the detected spectrum for the superradiant modes ($k_H<0$) is slightly different from the standard case. Since the roles of the creation and annihilation operators are interchanged for the superradiant modes, the expectation value of the number operator for the observers at $r_*\to \infty$ is computed as 
\begin{align}
    %\left\red{\langle} 
    \braket{a^{\rm up\dag}_{\omega lm} 
      a^{\rm up}_{\omega lm}}
      %\right\red{\rangle} 
   &= \frac{1}{-k_H}\sum_{{q}=1}^\infty ({q}+1) e^{{q}\beta k_H}\cr
   & = \frac{1}{-k_H}\frac{e^{-\beta k_H}}{e^{-\beta k_H}-1}\cr
   & = \frac{1}{k_H (e^{\beta k_H}-1)}. 
\end{align}
Although the path is different, the final result is exactly the same as in the usual case with $k_H>0$. For both positive and negative $k_H$, $I_\omega$ given in \eqref{Hawkingspectrum} remains positive. 
At the first sight, the spectrum of the Hawking radiation $I_\omega$ apparently diverges at $k_H=0$. 
However, from Eq.~\eqref{k3}, we find 
\begin{align}
    \Abs{A^{\rm out}}^{-2}=
    \frac{-k_H^2 \Abs{A^{\rm down}}^{-2}}{k_\infty({\cal J}k_\infty)}
    =\frac{k_H^2}{|k_\infty|^2} \Abs{A^{\rm down}}^{-2}.  
    \label{NandN}
\end{align}
Since $\Abs{A^{\rm down}}^2\ne 0$ at $k_H=0$ and $|k_\infty|^2\ne 0$ for $|\omega|>\mu$, Eq.~\eqref{NandN} implies that the spectrum of the Hawking radiation $I_\omega$ remains finite at $k_H=0$.  
In general, even for $k_\infty=0$, the ``up'' mode radial function simply becomes $\hat R^{\rm up}\to 1$ for $r_*\to\infty$, and is not singular. Therefore, $\hat R^{\rm up}$ is a non-vanishing finite real function, and $A^{\rm out}$ should not vanish. Hence, we expect $\Abs{A^{\rm out}}^{-2}$ remains to have the factor $k_H^2$ even for $k_\infty=0$.

\subsection{Problems with backreaction}
So far, we consider a scalar field on a fixed background Kerr spacetime. However, in a realistic situation the central black hole will spin down because of the extraction of angular momentum due to superradiance. In this section, we discuss how to describe the evolution of the system in such a time-dependent background. 

\subsubsection{Justification of the adiabatic evolution prescription}

Let us consider the superradiant discrete mode in a slowly varying background. For simplicity, we assume that only the spin parameter of the black hole changes adiabatically in time. In principle, our discussion can be extended to more general situations in which the black hole mass also evolves, or in which the effective mass of the field varies due to self-interactions. 

When the background evolution is adiabatic, {\it i.e.}, $\epsilon\equiv |{\nabla_\mu} \log a|/|\bar\omega|\ll 1$, we usually assume that the fastest-growing superradiant discrete mode with the complex frequency $\bar\omega$ continuously grows, even though the exact shape of the mode function and the frequency change in time. 
When the background becomes time-dependent, all that we need to do is to replace the mode functions with the ones that solve the KG equation on the given dynamical background, which is not so easy to perform analytically. 
However, when the background value of $a$ changes as a function of a time $\tau(t,r):=t - \Delta t(r)$, we can show that the mode function 
\begin{align}
u^{(0)}_{nlm}(x):=e^{-i\int^\tau d\tau' \bar\omega(\tau') +im\varphi} 
 \hat R^{(\tau)}_{nlm}(r) S_{\bar\omega(\tau) lm}(\theta), 
\end{align}
gives a sufficiently good approximation, where $\bar\omega(\tau):=\bar\omega(a(\tau))$ and 
\begin{align}
\hat R^{(\tau)}_{nlm}(r):= 
e^{- i\bar\omega(\tau) {\Delta t(r)}} R^{(a(\tau))}_{nlm}(r) ,
\end{align}
with $R^{(a)}_{nlm}(r)$ being the discrete radial mode function for a given spin parameter $a$. 
In $\hat R^{(\tau)}_{nlm}(r)$ the violent $\bar\omega$ dependence is assumed to be eliminated by the multiplicative phase factor $e^{-i\bar\omega(\tau) {\Delta t(r)}}$ for a large negative $r_*$, {which will be discussed in more details below.}  
{When we vary the background parameters, the label $n$ is assumed to be continuous, even though the order of growth rates may change. Namely,  $\bar\omega_{nlm}$ is supposed to have no discrete jump.}

To show that $u^{(0)}_{nlm}$ gives a good approximation, we consider an improvement of the solution by perturbatively adding corrections to $u^{(0)}_{nlm}$ around each reference time slice $\tau_*$ as 
\begin{align}
u_{nlm}=u^{(0)}_{nlm}+u^{(1,\tau_*)}_{nlm}+\cdots. 
\end{align}
We expand the K-G equation
\begin{equation}
   \left(\Box^{(a(\tau))}-\mu^2\right)u_{nlm}=0 ~ , 
\end{equation}
where $\Box^{(a(\tau))}$ is the d'Alembertian operator on the background with the variable $a(\tau(x))$.
To first order in $\epsilon$ at around $\tau=\tau_*$, {using the retarded Green's function $G^{(a)}_R(x,x')$ for a fixed value of $a$ (See Appendix~\ref{Sec:retardedGF}),} 
we find 
\begin{align}
u^{(1,\tau_*)}_{nlm}(x)=& \int d^4 x' \sqrt{-g(x')}G^{(a_*)}_R(x,x') {\cal S}(x'),
\end{align}
with $a_*:=a(\tau_*)$ and the source term is given by 
\begin{align}
{\cal S}(x):= 
\left(-\Box^{(a(\tau))}+\mu^2\right) u^{(0)}_{nlm}(x), 
\end{align}
{which should be of $\mathcal{O}(\epsilon)$ because it vanishes when $a$ is constant.}
Therefore, we find that the other modes are simply suppressed to be $O(\epsilon)$.

However, the contribution of the superradiant mode to the Green's function may secularly modify the evolution of the amplitude of the %\lyfu{fixed-background} 
{dominant mode, $u^{(0)}_{nlm}$}. However, as we shall see below, this variation is expected to be, at most, $O(\epsilon^2)$,  and hence it remains small of $O(\epsilon)$ even after the reaction time scale $O(\epsilon^{-1})$, as long as the gradient of the background is sufficiently small. 

First, we should notice that the source contains only the frequencies close enough to $\bar\omega^{(a_*)}_{nlm}$, as the frequency of the source ${\cal S}$ is limited to the modulation of the seed frequency $\bar\omega^{(a_*)}_{nlm}$ caused by low frequency perturbations. 
In general, we can expand $u_{nlm}$ in terms of a complete set of mode functions, $\{u^{(a_*)}_{nlm},u^{(a_*)}_\Lambda\}$ for $a=a_*$, where $\Lambda$ represents the labels of all modes with the frequency close to $\bar\omega_{nlm}$, excluding the seed mode labeled by $nlm$ itself. Here, we associate the superscript ``$(a)$'' to specify the modes on the background with $a=a_*$.

Formally, we expand $u_{nlm}$ as $u_{nlm}\approx c_{nlm}(\tau) u^{(a_*)}_{nlm}+\sum_\Lambda c_\Lambda(\tau) u^{(a_*)}_\Lambda$ on the time slice $\tau=\tau_*$ with slowly changing coefficients depending on $\tau$. 
By construction, we can choose $c_{nlm}(\tau_*)=1$ as the initial state.  
We can also construct ${\cal J} u_{nlm}$ in the same way. 
Then, from the conservation of the KG inner product {$\lround u_{nlm},\mathcal J u_{nlm}\rround$}, we 
find $\Abs{c_{nlm}}^2=1-\sum_\Lambda|c_\Lambda|^2$. {As $c_\Lambda$ is sourced by $\cal S={\cal O}(\epsilon)$, we have $c_\Lambda \sim \mathcal{O}(\epsilon)$.}
This implies $(c_{nlm}-1) + ({\cal J}c_{nlm}-1)=O(\epsilon^2)$. 
However, we may worry if
each of $c_{nlm}$ and $({\cal J}c_{nlm})$ may change. Namely, the question is whether $dc_{nlm}(\tau)/d\tau|_{\tau=\tau_*}=O(\epsilon^2)$ or not. As the amplitude of $u_{nlm}$ is increasing in time, one can always tune the origin of the time coordinate such that $c_{nlm}(\tau_*)=1$, as mentioned above. 

To examine the change rate of $c_{nlm}(\tau)$, we focus on the ``particle number'' defined by the integral of $n^{(\tau)}_\mu j^\mu(u_{nlm},u^*_{nlm})$ in some finite region between $r_{\rm min}$ and $r_{\rm max}$, {where $n^{(\tau)}_{\mu}$ is the unit vector normal to the constant-$\tau$ time-slice and $j^\mu$ is defined in Eq.~\eqref{def:j}.}
In the adiabatic evolution, the evolution of the overall amplitude corresponding to $c_{nlm}$ is determined by the balance between the change rate of this particle number and the number fluxes.

We can estimate the number flux at the boundaries at $r_{\rm min}$ and $r_{\rm max}$.  
{The particle number flux differs from the case with fixed $a$ primarily because of two effects.}
One is the flux due to the extra modes $\{u_\Lambda\}$ and the other is the correction to the flux carried by $u^{(a_*)}_{nlm}$ caused by the adiabatic change of $a$. 
The former is manifestly $O(\epsilon^2)$ because the flux is quadratic in the amplitude of modes, which is $O(\epsilon)$.  
The latter is also $O(\epsilon^2)$ because this correction is $O(\epsilon)$ higher-order correction to the original flux. However, the original flux carried by $u^{(a_*)}_{nlm}$ is already $O(\epsilon)$, since the smallness of this flux is the origin of the adiabatically slow evolution of $a$. Thus, we find that the correction is at most $O(\epsilon^2)$. 
Therefore, we can conclude $d|c_{nlm}(\tau)|^2/d\tau|_{\tau=\tau_*}=O(\epsilon^2)$, or equivalently, $dc_{nlm}(\tau)/d\tau|_{\tau=\tau_*}=O(\epsilon^2)$,
and the correction to the adiabatic evolution of the number of particles contained in the cloud is suppressed by $\epsilon$, which justifies the use of the adiabatic approximation.

In the above discussions, we assumed that {the variation of $u^{(0)}_{nlm}(x)$ through $\bar\omega(\tau)$, {\it i.e.}, $\displaystyle \frac{du^{(0)}_{nlm}(x)}{d\bar\omega(\tau)}\frac{d\bar\omega(\tau)}{d\tau}$ is small $\propto \epsilon$.} 
This condition might be violated at a large negative $r_*$.   
Extracting the $\tau$-dependence in this region, we find 
\begin{align}
u^{(0)}_{nlm}\propto 
  e^{-i\int^\tau d\tau' \bar\omega(\tau') - i\bar\omega(\tau)(\Delta t(r) + r_*)}. \nonumber
\end{align}
Taking its $\tau$-derivative thorough $\bar\omega(\tau)$, we have 
\begin{align}
 \frac{d \log u^{(0)}_{nlm}}{d\bar\omega(\tau)}\frac{d\bar\omega(\tau)}{d\tau}
  =&
%i\bar\omega(\tau)\left(1-\frac{dt(r,\tau)}{d\tau}\right)\cr 
- i\frac{d\bar\omega(\tau)}{d\tau} \left(\Delta t(r) + r_* \right). 
\end{align}
For both terms to be small, we should choose $\Delta t(r)= - r_*$ in this region, {\it i.e.}, the time slicing should be along the outgoing null coordinate in the near horizon region.  
This is not so demanding, because all the fluxes are pointing toward the horizon, and the propagation speed gets close to the speed of light there. 

\subsubsection{Vanishing of a superradiant mode}
As we change the parameters of the system like $a$ and $\mu$, the superradiance frequency $\bar\omega_{nlm}$ varies. 
There will be a boundary of the parameter space at which a sequence of superradiant modes stops existing. 
Here, we discuss what happens at this boundary. 

The boundary will be characterized by $\bar\omega_I:={\rm Im} \,\bar\omega_{nlm}=0$. 
As we have shown in Sec.~\ref{sec:growth_of_superrad}, superradiant poles specified by $\bar\omega_{nlm}$ can exist only in the range \eqref{Cond:SRcomp}. 
Therefore, when $\bar\omega_I$ gets close to 0, we should have 
\begin{align}
    \bar\omega_R(m \Omega_H-\bar\omega_R)=-\bar\omega_R k_H(\bar\omega_R,m)>0,  
\end{align}
where $\bar\omega_R:={\rm Re} \,  \bar\omega$. 
In fact, we can prove that at the boundary the above condition is saturated, {\it i.e.}, $k_H(\bar\omega_R,m)=0$, if $\bar\omega_R<\mu$ at the boundary. 
Recall that the superradiant mode is given by an analytic continuation of $u^{\rm up}_{\omega lm}$ off the real axis. 
As long as the region of $\omega$ that does not cross the branch cut, the exponential decaying condition is guaranteed, except on the real axis. 
If $\bar\omega_R<\mu$ in the $\bar\omega_I\to 0$ limit, 
the un-normalized radial function
\begin{align}
\hat R^{\rm up}_{nlm}:=R^{\rm up}_{nlm}/{\cal N}^{\rm up}_{nlm},
\end{align}
becomes real in this limit, {i.e.}, $A^{\rm in}=A^{{\rm out}*}$. 
Since $A^{\rm out}_{\bar\omega lm}=0$ according to the definition of a discrete mode, 
discrete modes cannot exist, as long as we can define $A^{\rm in}$ and $A^{\rm out}$. 
However, the distinction between ``in''-going waves and ``out''-going waves becomes unclear for $k_H=0$.  
Therefore, we conclude that $k_H=0$ should be satisfied when a discrete mode crosses the real axis.  
The same conclusion follows from the implication that the contribution of the mode of our interest to the energy flux evaluated in Eq.~\eqref{eq:positiveFlux} should vanish for $\bar\omega_I\to 0$.  

On the other hand, if $\bar\omega$ hits the boundary of $\bar\omega_I =0$ with $\bar\omega_R>\mu$ 
, the un-normalized ``up"-type radial function can have an unsuppressed flux to infinity, and hence $\omega_I=0$ can be realized because of the balance between the fluxes from the horizon and to infinity. As a result, we cannot specify the value of $\bar\omega_R$ at $\omega_I=0$ from the formal discussion.
In this case, the boundary is on the branch cut. 
From the continuity, it is expected that the sequence continues to the next Riemann surface, {in which ${\rm Im} \, k_\infty<0$. 
We are not completely sure, but this case may not happen. The superradiant modes with a lower frequency $\omega_{\rm R}$ and a faster growth rate, which are usually of our interest, always cross $\omega_{\rm I}=0$ with $\omega_{\rm R} < \mu$.}
{In both cases, after continuously changing the parameters further beyond the boundary values, the mode should become a QNM.} 

In the latter case, the flux evaluated near the horizon \eqref{eq:positiveFlux} does not vanish in the limit $\omega_I\to 0$. 
This flux is to be compensated by the energy flux at a large positive $r_*$. 
We can evaluate the flux in the same way as before, to obtain 
\begin{align}
\bra{B} T_{\mu\nu} \ket{B}&  \hat r^\mu \xi^\nu\cr
\to &\sum_{nl}\sum_{m>1}^l \left({\cal N}_{\bar\omega lm}^{\rm DSR}\right)^2 
    (\bar\omega k_\infty^*(\bar\omega)+
      \bar\omega^* k_\infty(\bar\omega))  \cr
 & \qquad 
   \times e^{-2({\rm Im} k_\infty) r_*} 
      \cosh\left(2 \bar\omega_I t\right), 
\label{eq:positiveFluxatlarger}
\end{align}
at a large $r_*$ , and we can confirm that the propagation speed of the energy flux, which we define by the slope of the constant energy flux surface, is given by 
\begin{align}
 \pm\frac{dr_*}{dt} = v_{nlm} &:=\frac{\bar\omega_I}{{\rm Im}\, k_\infty(\bar\omega)}\cr
 & =\frac{{\rm Re}\, k_\infty(\bar\omega)}{\bar\omega_R}\approx
 \left.\frac{d\omega}{dk_\infty}\right|_{\omega=\bar\omega_R},  
\end{align}
which approximately agrees with the group velocity for $\bar\omega_I\ll \bar\omega_R$. 
In the last approximate equality, we assumed $\bar\omega_I\ll \bar\omega_R$. 
We find that the flux given in Eq.~\eqref{eq:positiveFluxatlarger} cancels the flux evaluated in Eq.~\eqref{eq:positiveFlux} at $\omega_I=0$, using the relation $(A^{\rm in})^2=-k_\infty/k_H$ obtained from Eq.~\eqref{k1} with the aid of Eq.~\eqref{JArelation}.

\subsection{The direction of level transitions of an interacting field}
We consider an interacting field in this section. 
For concreteness, we consider $\lambda\phi^4$ interaction, 
in which 2-2 scattering process is possible. 
We consider two processes as examples. The classical treatment of scattering induced by the self-interaction has been studied in various literature, see for example~\cite{Baryakhtar:2020gao, Omiya:2022gwu, Witte:2024drg}.

\subsubsection{Decay of the secondary cloud}
 In the first example, we assume that the discrete mode with $(n,l,m)=(2,1,1)$ is the fastest growing mode and forms a dominant component of the cloud. We consider the process in which two particles in this dominant mode make a transition to a particle labeled as $(n,1,1)$ with $n \ge 3$ and an ``up'' mode particle with $\omega=\omega_c:={\rm Re}\,(2\omega_{211}-\omega_{n11})$ and $m=1$.  Note that $\omega_c < \mu$ and thus no flux to infinity, i.e., "in" modes are not involved.
In this case, the actual direction of the reaction turns out to be opposite, and as a result, the growth of the secondary cloud is suppressed by the interaction.  

It is convenient to use the mode function introduced in Eq.~\eqref{temporalSRModes} and the associated creation and annihilation operators, $\tilde a_{nlm}^\dag$ and $\tilde a_{nlm}$.
Picking up the term related to this process at around $t=t_0$, the interaction Hamiltonian is schematically written as 
\begin{align}
  H_{\rm int} = {\cal O}+{\cal O}^\dag+\cdots,
\end{align}
with
\begin{align}
  {\cal O}=\sum_l {\cal A}_l\,
  \tilde a^{(t_0)}_{n11}
  \left(\tilde  a^{(t_0)\dag}_{211}\right)^2 
  a^{({\rm up})\dag}_{\omega_c l1}, 
\end{align}
where ${\cal A}_l$ represents the amplitude for the process concerning the mode specified by the label $(\omega_c,l,1)$, which we treat as if a discrete mode for brevity. 
In the following, we consider a single $l$-mode, for simplicity. 
Here, the point is that the continuous ``up'' mode particle involved in the above reaction is superradiant, and hence the creation and annihilation operators are interchanged as discussed in Sec.~\ref{Sec.Hamiltonian}. 
Then, the probability of the forward process {described by the operator ${\cal O}$ composed of 
one annihilation and three creation operators}, which can be interpreted as the decay of the particle labeled with $(n,1,1)$ into three particles, two $(2,1,1)$ and one $(\omega_c, l,1)$, is proportional to 
\begin{align}
 \langle {\cal O}^\dag {\cal O} \rangle
 \propto \tilde N^{(t_0)}_{n11} 
    (\tilde N^{(t_0)}_{211}+1)
    (\tilde N^{(t_0)}_{211}+2)
    \braket{a^{(\rm up)}_{\omega_c l m} 
    a^{({\rm up})\dag}_{\omega_c l m}}. 
\end{align}
On the other hand, the probability of the inverse process is given by 
\begin{align}
 \langle {\cal O} {\cal O}^\dag \rangle
 \propto (\tilde N^{(t_0)}_{n11}+1) 
    \tilde N^{(t_0)}_{211}(\tilde N^{(t_0)}_{211}-1)
    \braket{
    a^{({\rm up})\dag}_{\omega_c l m}
    a^{(\rm up)}_{\omega_c l m}}. 
\end{align}

For low energy ``up'' modes with $\omega<\mu$, it would be appropriate to take the thermal state, and the occupation number of the dominant mode should be large. Therefore we have 
\begin{align}
& \langle {\cal O}^\dag {\cal O} \rangle
 \propto \tilde N^{(t_0)}_{n11} 
    (\tilde N^{(t_0)}_{211})^2
    \frac{e^{\beta |k_H(\omega_c,1)|}}
      {e^{\beta |k_H(\omega_c,1)|}-1},\cr
& \langle {\cal O} {\cal O}^\dag \rangle
 \propto (\tilde N^{(t_0)}_{n11}+1) 
    (\tilde N^{(t_0)}_{211})^2
    \frac{1}
      {e^{\beta |k_H(\omega_c,1)|}-1}. 
\end{align}
Unless $\tilde N^{(t_0)}$ is very close to 0, the forward process, {\it i.e.}, the decay of a particle in the secondary cloud in the state $(n,1,1)$, dominates the inverse process. 
We also find that the two processes balance at 
\begin{align}
  \tilde N^{(t_0)}_{n11}= \frac{1}
      {e^{\beta |k_H(\omega_c,m)|}-1}, 
\end{align}
which is not a large number, compared to the particle number of the macroscopic axion cloud one considered in the literature. 

\subsubsection{Decay of the primary cloud}

The situation is different when we consider higher $m$-mode as the secondary cloud. In this case, an example of the most relevant interaction would be 
\begin{align}
  {\cal O}'=\,
  \left(\tilde  a^{(t_0)}_{211}\right)^2 
  \tilde a^{(t_0)\dag}_{322}
  a^{({\rm up})\dag}_{\omega'_c 00}, 
\end{align}
with $\omega'_c:={\rm Re}\,(2\omega_{211}-\omega_{322})$. 
The operator ${\cal O}'$ represents the process of the decay of two particles in the dominant cloud ($(2,1,1)$ state) into a particle in the secondary cloud ($(3,2,2)$ state) and that in the continuous non-superradiant ``up'' mode with $m=0$. 
In this case, the probabilities of the forward process and the reverse process are, respectively,  given by
\begin{align}
& \braket{ {\cal O}^\dag {\cal O} }
 \propto (\tilde N^{(t_0)}_{322}+1) 
    (\tilde N^{(t_0)}_{211})^2
    \frac{e^{\beta |k_H(\omega'_c,0)|}}
      {e^{\beta |k_H(\omega'_c,0)|}-1}, \cr
& \braket{{\cal O} {\cal O}^\dag}
 \propto \tilde N^{(t_0)}_{322} 
    (\tilde N^{(t_0)}_{211})^2
   \frac{1}
      {e^{\beta |k_H(\omega'_c,0)|}-1}. 
\end{align}
The net transition rate is determined by the difference between the forward process and the reverse process, {\it i.e.}, $\propto \lround {\cal O}^\dag {\cal O} \rround -\lround {\cal O} {\cal O}^\dag \rround$. 
Therefore, when the occupation number of the secondary cloud is large enough, the factor depending on the Hawking temperature drops off in the net transition rate, which justifies the computation of the transition rate based on the classical energy flux toward the horizon.

\section{Conclusion}
\label{Sec.Conclusion}
Superradiant instability has been extensively studied as a classical phenomenon, but a clear quantum interpretation has long been lacking. In this paper, we developed a quantum framework of superradiance by performing the canonical quantization of a massive real scalar field around a Kerr black hole, including both continuous and discrete modes, which are normalizable. 

For the discrete modes corresponding to those in the superradiant clouds, the standard normalization fails because the inner product $\lround u_{nlm}, u^*_{nlm} \rround$ vanishes (otherwise the inner-product necessarily becomes time-dependent). This orthogonality is due to the cancellation between the contributions from the inside and outside of the potential barrier. Instead of the ordinary complex conjugation, the conjugate mode having non-vanishing inner-product with $u_{nlm}$ is obtained via the operation $\mathcal{J}$, which flips the sign of $(\omega, m)$ to $(-\omega, -m)$. This choice of a complete orthonormal set leads to a commutation relation differing from the standard one after canonical quantization (See Eq.~\eqref{nonstarndardCR}.). 
To construct a pair of positive- and negative-frequency functions that are complex conjugate to each other, we relax the requirement that the mode function be the eigenfunction of time translation, and introduce new modes $v_{\pm}$ as linear combinations of $u_{nlm}$ and $\mathcal{J}u_{nlm}$ among the discrete modes. In this construction, $v_{+}$ is the one that has more weight on the outside of the potential barrier, while $v_{-}$ is opposite.

After performing the mode expansion, we derive explicit expressions for the Hamiltonian and angular momentum. We consider a ``Boulware-like'' stationary reference state, albeit finding that no truly stationary state exists for the discrete modes. We then study the time evolution of these discrete modes, which we expect to represent superradiant clouds. The growth is attributed to the time dependence of the mode function itself. The norm of the functions $v_\pm$ is conserved, although they grow in time. This conservation of norm may look contradictory, but it can be explained by the negative contribution of the conserved number density inside the potential barrier. The constancy of the norm simply results from a cancellation between the increase in the positive particle number outside the barrier and the negative particle number inside. To capture the growth of the superradiant cloud, we introduced a quasi-particle number that counts only the portion of the number density outside the potential barrier. Using this definition, we have shown that this quasi-particle number grows exponentially in time, regardless of the choice of initial state.

We also explored several related topics. For the continuous modes, we considered the Unruh vacuum state, and calculated the spectrum of Hawking radiation. If we naively apply the Bose-Einstein statistical factor, the flux to infinity seems to become negative for modes whose energy measured by a local observer near the horizon is negative. However, such a pathology is evaded by the flip of the sign in the grey body factor, explaining why the energy flux remains positive. Furthermore, we discuss how the backreaction on the black hole spin influences the evolution of superradiant clouds. We show that it is appropriate to evolve the cloud adiabatically, based on the flux balance with respect to the particle number. We also clarify what occurs when the spin decreases to the saturation value, at which point the superradiant mode transitions into a quasi-normal mode (QNM). Finally, we give a quantum mechanical explanation to the direction of level transitions that arises in the presence of self-interactions.

Our paper raised a number of interesting follow-up questions. For example, as mentioned earlier, how to choose a physically realistic initial state corresponding to the state after the gravitational collapse. A complete quantum mechanical analysis should be formulated in a dynamical spacetime describing BH formation through collapse, which has not been studied before. Moreover, the quantization procedure can be extended to other types of fields, such as free fermions and vectors, where new issues or phenomena may arise.

%\textcolor{red}{=====================}

%%%%%%%%%%%%%%%%%%%%%%%%%%%%%%%%%%%%%%%%%%%%%%%

\section{Acknowledgment}

We thank Ali Akil, Andrew Cohen and Jiarong Zhang for inspiring discussions. H. O. is supported by JSPS KAKENHI Grant Numbers JP23H00110 and Yamada Science Foundation. T. T. is supported by JSPS KAKENHI Grant Nos. JP23H00110, JP24H00963, and JP24H01809. X.T is supported by STFC consolidated grants ST/T000694/1 and ST/X000664/1. Y.W. is supported by the National Key R\&D Program of China (2021YFC2203100), and GRF grant 16306422 by the RGC of Hong Kong SAR. H.Y.Z. is supported by IBS under the project code, IBS-R018-D3. 

\appendix

\section{Normalization of Discrete Modes}
\label{Appendix:DiscreteNorm}
Here, we discuss the normalization of the discrete modes. We evaluate Eq.~\eqref{Eq.Normu}. 
For this purpose, we evaluate $\eta$ at $\omega=\bar\omega_{nlm}$, using the equation of motion for $S_{nlm}(\theta)$. 
Taking the derivative of Eq.~\eqref{polar equation} with respect to $\omega$, we have 
\begin{align}
 \int d(\cos\theta) S(\theta) \biggl[
 &\left(2am 
 - 2a^2\omega \sin^2\!\theta
 +\frac{d\lambda}{d\omega} 
   \right) S(\theta) \cr
& \qquad +{\cal L}_\theta \frac{dS(\theta)}{d\omega} 
 \biggr] = 0 ~ .
\end{align} 
The term with $dS(\theta)/d\omega$ is removed by performing integration by parts, owing to self-adjoint nature of ${\cal L}_\theta$ and the regularity of $S(\theta)$ on both boundaries. 
From this identity, we obtain 
\begin{align}
  \eta=\frac{m}{a\omega}+\frac{1}{2a^2\omega}
  \frac{d\lambda}{d\omega}.
\end{align}
Substituting this identity into $f(r)$ given in Eq.~\eqref{Eq:fofr}, we obtain
\begin{align}
        f(r) = 2\omega- \frac{{2}am}{r^2 + a^2} - \frac{\Delta}{(r^2 + a^2)^2}
        \frac{d\lambda}{d\omega} ~ .
\end{align}

Denoting $R_{nlm}(r)$ by using the un-normalized mode function $R^{\rm up}_{\omega lm}(r)$ as $R_{nlm}(r)=
{\cal N}^{\rm DSR}_{nlm}
\hat R^{\rm up}_{\bar\omega_{nlm} lm}(r)$. 
Then, we take the derivative of Eq.~\eqref{radial equation} in the same way, to obtain
\begin{align}
 \int dr_* &\frac{f(r)}{{\cal G}^2}R_{nlm}^2\cr
 & =-({\cal N}^{\rm DSR}_{nlm})^2
   \int dr \hat R^{\rm up}_{nlm} {\cal L}_r 
  \left.\frac{d\hat R^{\rm up}_{\omega lm}}{d\omega}\right\vert_{\omega=\bar\omega_{nlm}}
  \cr
 & =\Biggl[ \frac{({\cal N}^{\rm DSR}_{nlm})^2}{{\cal G}^2}
  \biggl(
  \frac{d\hat R^{\rm up}_{\omega lm}}{dr_*}
   \frac{d\hat R^{\rm up}_{\omega lm}}{d\omega}\cr
&\qquad\quad    -\hat R^{\rm up}_{\omega lm} 
  \frac{d^2\hat R^{\rm up}_{\omega lm}}{dr_* d\omega}\biggl)_{\omega=\bar\omega_{nlm}}\Biggr]^{{r_* = \infty}}_{{r_*=-\infty}}~. 
\end{align}
Substituting Eq.~\eqref{DefRup} and 
\begin{align}
  &\left. \frac{d R_{\omega lm}(r)}{d\omega}\right\vert_{\omega=\bar\omega_{nlm}}\cr
  & \quad
  \to{\cal N}^{\rm DSR}_{nlm}
       {\mathcal G}
       \left\{
  \begin{array}{ll}
   \displaystyle\frac{dA_{\omega l m}^{\rm out}}{d\omega} e^{ik_Hr_*}+(\cdots)e^{-ik_Hr_*},  
   \hspace*{-2cm} 
   &\cr
   & (r_* \to -\infty),\cr
        e^{+ ik_\infty r_*},  & (r_* \to +\infty), 
        	\end{array}\right.
            \label{Eq.dRQNM}
\end{align}
we find
\begin{align}
 \int dr_* &\frac{f(r)}{{\cal G}^2}R_{nlm}^2
 =\left. 2i({\cal N}^{\rm DSR}_{nlm})^2 k_H A^{\rm in}_{\omega lm}\frac{dA^{\rm out}_{\omega lm}}{d\omega}\right\vert_{\omega=\bar\omega_{nlm}}. 
\end{align}
Using the first equality in Eq.~\eqref{JArelation}, we arrive at Eq.~\eqref{discrete_norm}. 

\section{Retarded Green's function}
\label{Sec:retardedGF}
In quantum theory, we often use the retarded Green's function given by 
\begin{align}
  G_R(x,x')=i\theta(t-t')[\Phi(x),\Phi(x')],  
  \label{Def:Greenfn1}
\end{align}
which is manifestly 0 when $t<t'$. 
It is easy to show that this Green's function satisfies
\begin{align}
 \left(\Box -\mu^2\right)G_R(x,x')=
 -\frac{\delta^4(x-x')}{\sqrt{-g(x)}}. 
\label{Eq:Greenfn}
\end{align}
By using the unit vector ${n}^\mu$ normal to a constant-$t$ time slice $\Sigma_t$, 
\begin{align}
\Box \Phi=\frac1{\sqrt{-g}}\partial_\mu \left\{\sqrt{-g}
 \left(-{n}^\mu {n}^\nu+ h^{\mu\nu}\right)
 \partial_\nu \Phi\right\}. 
\end{align}
Since $h^{\mu\nu} {n}_\nu=0$, $\partial_\mu\sqrt{-g} h^{\mu\nu}\partial_\nu$ do not contain time-derivative. The time derivative is only in 
$\partial_\mu\sqrt{-g} {n}^{\mu} {n}^{\nu}\partial_\nu$. One time derivative should act on $\theta(t-t')$, but the contribution with the first derivative acting on $\theta(t-t')$ vanish because of the equal time commutation relation, $[\Phi(t,\bx),\Phi(t,\bx')]=0$. So, the non-trivial contribution remains only when the second derivative acting on $G_R$ operates on $\theta(t-t')$.  
Recalling that the conjugate momentum is defined by $\Pi(x):={n}^t {n}^\mu\partial_\mu \Phi(x)$, we have 
\begin{align}
\left(\Box -\mu^2\right)G_R(x,x')=-i\delta (t-t')[\Pi(t,\bx),\Phi(t,\bx')]. 
\end{align}
Using the commutation relation at the same time, $[\Pi(t,\bx),\Phi(t,\bx')]=-i\delta^3(\bx-\bx')/\sqrt{-g}$, we obtain Eq.~\eqref{Eq:Greenfn}.

More explicitly, substituting the mode decomposition of 
$\Phi(x)$, we find 
\begin{align}
G_R&(x,x')\cr
=&i\theta(t-t') 
\sum_{lm}\Biggl\{
\int_{-\infty}^\infty \frac{d\omega}{k_H(\omega,m)} 
u^{\rm up}_{\omega lm}(x) 
({\cal J}^{\Re}  u^{\rm up}_{\omega lm}(x'))\cr
& +\left(\int_{-\infty}^{-\mu}+\int_\mu^\infty\right)
 \frac{d\omega}{k_\infty(\omega)} u^{\rm in}_{\omega lm} (x)
 ({\cal J}^{\Re}  u^{\rm in}_{\omega lm}(x'))\cr
&+\sum_n u_{nlm}(x) ({\cal J}u_{nlm}(x'))
  +({\cal J}u^*_{nlm}(x)) u^*_{nlm}(x') 
 \Biggr\}\cr
=& i\theta(t-t') \sum_{lm}\Biggl[\Biggl(
\int_{-\infty}^\infty d\omega
\frac{\hat R^{\rm up}_{\omega lm}(r) 
({\cal J}^{\Re}  \hat R^{\rm up}_{\omega lm}(r'))}{4\pi k_H(\omega,m)\Abs{A^{\rm out}_{\omega lm}}^2} \cr
& +\left(\int_{-\infty}^{-\mu}+\int_\mu^\infty\right)
 d\omega 
 \frac{k_\infty(\omega)\hat R^{\rm in}_{\omega lm} (r)
 ({\cal J}^{\Re}  \hat R^{\rm in}_{\omega lm}(r'))}{4\pi k^2_H(\omega,m)\Abs{A^{\rm out}_{\omega lm}}^2}\Biggr)\cr
& \quad \times e^{-i\omega(t-t')+im(\varphi-\varphi')} 
  S_{\omega lm}(\theta)S_{\omega lm}(\theta')\cr
& +\sum_n \bigl( R_{nlm}(r) R_{nlm}(r') S_{\bar\omega_{nlm} lm}(\theta)S_{\bar\omega_{nlm} lm}(\theta')\cr
& \quad \times  e^{-i\bar\omega_{nlm}(t-t')}+R^*_{nlm}(r) R^*_{nlm}(r')
S_{\bar\omega^*_{nlm} lm}(\theta)\cr
& \quad \times S_{\bar\omega^*_{nlm} lm}(\theta')
  e^{-i\bar\omega^*_{nlm}(t-t')}
  \bigr) e^{im(\varphi-\varphi')} 
 \Biggr].
 \label{Eq:Green1}
\end{align}

On the other hand, we can construct the retarded Green's function assuming the decomposition into the Fourier modes and Harmonics in the three dimensional time-like surface spanned by $(t,\theta,\varphi)$ as
\begin{align}
\bar G_R(x,x')=
& \sum_{lm}\int_C \frac{d\omega}{2\pi} e^{-i\omega(t-t')} 
G_{\omega lm}(r,r')\cr
& \quad \times e^{im(\varphi-\varphi')} 
  S_{\omega lm}(\theta)S_{\omega lm}(\theta'), 
  \label{Eq:Green2}
\end{align}
where the integration contour $C$ is chosen to run from $-\infty$ to $+\infty$ in the complex plane, lying entirely above all $\bar\omega_{nlm}$, {\it i.e.}, the zeros of $A^{\rm out}_{\omega lm}$.
If $G_{\omega lm}(r,r')$ satisfies 
\begin{align}
 {\cal L}_r G_{\omega lm}(r,r') =-\delta(r-r'), 
 \label{Eq:Gr}
\end{align}
since $\Box =(\sin\theta/\sqrt{-g}){\cal L}_r+\cdots$, we find 
\begin{align}
 \left(\Box-\mu^2\right)&\bar G_R(x,x')=
 -\frac{\sin\theta}{\sqrt{-g}}\delta(r-r')
  \sum_{lm}\int_C \frac{d\omega}{2\pi}\cr
 & \times e^{-i\omega(t-t')+im(\varphi-\varphi')} 
  S_{\omega lm}(\theta)S_{\omega lm}(\theta')\cr
&\qquad \qquad =-\frac{\delta^4(x-x')}{\sqrt{-g(x)}}. 
\end{align}
We can construct $G_{\omega lm}(r,r')$ that satisfies Eq.~\eqref{Eq:Gr} by choosing 
\begin{align}
  G_{\omega lm}(r,r')= &
   \frac{i}{2k_H A^{\rm out}_{\omega lm}}\biggl(
\hat R^{\rm up}_{\omega lm}(r)\hat R^{\rm in}_{\omega lm}(r')\theta(r-r')\hspace{-6mm}\cr
& \quad    +\hat R^{\rm up}_{\omega lm}(r')\hat R^{\rm in}_{\omega lm}(r)\theta(r'-r)
     \biggr), 
\end{align}
where we use $W[\hat R^{\rm up},\hat R^{\rm in}]=-2ik_H A^{\rm out}_{\omega lm}$.
From the choice of the contour $C$, we can easily show that $\bar G(x,x')=0$ for 
$t<t'$ by closing the integration in the upper half complex plane. 

Now, the problem is to show the equivalence between the expressions given in \eqref{Eq:Green1} and \eqref{Eq:Green2}. For this purpose, we consider the case in which $t>t'$ and $r>r'$. The equivalence in the case with $r<r'$ can be shown in the same way. 
We start with rewriting $G_R(x,x')$ given in Eq.~\eqref{Eq:Green1}. 
The term with $\hat R^{\rm in}_{\omega lm} (r)
 ({\cal J}^{\Re}  \hat R^{\rm in}_{\omega lm}(r'))$ is rewritten by using 
 $\int_{-\infty}^{-\mu} d\omega (\cdots )= \int_{\mu}^{\infty} d\omega {\cal J}^{\Re}(\cdots )$ as 
\begin{align}
 -i\sum_{lm}
 \left(\int_{-\infty}^{-\mu}+\int_\mu^\infty\right)
 d\omega 
 \frac{k_\infty(\omega)
 ({\cal J}^{\Re} \hat R^{\rm in}_{\omega lm} (r))
  \hat R^{\rm in}_{\omega lm}(r')}
  {4\pi k^2_H(\omega,m)\Abs{A^{\rm out}_{\omega lm}}^2}. 
\label{Eq:SecondTermGreen}
\end{align}
Using the following identities,
\begin{align}
 {\cal J}^{\Re} \hat R_{\omega lm}^{\rm up}(r)
&=({\cal J}^{\Re} A^{\rm in}_{\omega lm})
 ({\cal J}^{\Re} \hat R^{\rm in}_{\omega lm}(r))\cr
&\qquad\qquad +({\cal J}^{\Re} A^{\rm out}_{\omega lm})
 (\hat R^{\rm in}_{\omega lm}(r)),
 \label{Eq:IRup}
 \\
  {\cal J}^{\Re} \hat R_{\omega lm}^{\rm in}(r)
&=({\cal J}^{\Re} A^{\rm up}_{\omega lm})
 ({\cal J}^{\Re} \hat R^{\rm up}_{\omega lm}(r))\cr
&\qquad\qquad +({\cal J}^{\Re} A^{\rm down}_{\omega lm})
 (\hat R^{\rm up}_{\omega lm}(r))\cr
&=\frac{k_H(\omega,m)}{k_\infty(\omega)}
  \biggr[-A^{\rm in}_{\omega lm}
 ({\cal J}^{\Re} \hat R^{\rm up}_{\omega lm}(r))\cr
&\qquad\qquad
 +({\cal J}^{\Re} A^{\rm out}_{\omega lm})
 (\hat R^{\rm up}_{\omega lm}(r))\biggr],  
 \label{Eq:IRin}
\end{align}
with 
\begin{align}
  {\cal J}^{\Re} A^{\rm in}_{\omega lm}=
  \left\{\begin{array}{ll}
  A^{\rm out}_{\omega lm}, & (|\omega|<\mu),\cr
  \displaystyle
  \frac{k_\infty(\omega)}{k_H(\omega, m)}
     A^{\rm up}_{\omega lm}, & (|\omega|>\mu),
     \end{array}
  \right.
\end{align}
the whole contribution of the continuous modes for $t>t'$ becomes 
\begin{align}
G_{R}^{\rm conti}&(x,x')
\cr=&
\frac{i}{4\pi}
 \sum_{lm} \Biggl[ \int_{-\infty}^{\infty}
\hat R^{\rm up}_{\omega lm}(r) 
 ({\cal J}^{\Re} A^{\rm out})\hat R^{\rm in}_{\omega lm}(r')\cr
 & + \int_{-\mu}^{\mu}
  A^{\rm out} \hat R^{\rm up}_{\omega lm}(r) 
({\cal J}^{\Re} \hat R^{\rm in}_{\omega lm}(r'))
 \cr
& {-}\left(\int_{-\infty}^{-\mu}+\int_\mu^\infty\right)
({\cal J}^{\Re} A^{\rm out}) \hat R^{\rm up}_{\omega lm}(r) \cr
&\qquad \qquad \times 
 \hat R^{\rm in}_{\omega lm}(r'))
 \Biggr]
 \frac{d\omega}{k_H \Abs{A^{\rm out}}^2}\cr
 & ~~ \times e^{-i\omega(t-t')+im(\varphi-\varphi')} 
  S_{\omega lm}(\theta)S_{\omega lm}(\theta'). 
 \label{Eq:GRconti}
\end{align}
In this calculation, the contribution coming from the first term in \eqref{Eq:IRup} and 
and the first term in \eqref{Eq:IRin} mutually cancel in the region $(-\infty,-\mu)$ 
and $(\mu,\infty)$. 

Here, one can prove  
\begin{align}
&A^{\rm out}={\cal J}^{\Re} A^{\rm in},
\quad \Abs{A^{\rm out}}^2=\Abs{A^{\rm in}}^2,\cr
& {\cal J}^{\Re} \hat R^{\rm in}_{\omega lm}=
     {\cal J} \hat R^{\rm in}_{\omega lm},
\qquad (\mbox{for } |\omega|<\mu), 
\end{align}
which we can apply to the second term in the square brackets of \eqref{Eq:GRconti}.  
We flip the sign of the integration variable $\omega$ in the third term, 
by applying 
$\int_{-\infty}^{-\mu} d\omega (\cdots )= \int_{\mu-i\epsilon}^{\infty-i\epsilon} d\omega {\cal J}(\cdots )$, and using 
\begin{align}
& {\cal J} {\cal J}^{\Re} A^{\rm out}={\cal J}^{\Re} A^{\rm in}, 
 {\cal J}\Abs{A^{\rm out}}^2=\Abs{A^{\rm in}}^2,
 \cr
&  {\cal J} \hat R^{\rm up}_{\omega lm}=\hat R^{\rm up}_{\omega lm},
\end{align}
we find 
\begin{align}
&\mbox{``The second and the third terms} \cr
&\qquad \mbox{in the square brackets of Eq.~\eqref{Eq:GRconti}''}\cr
&=\frac{i}{4\pi}
 \sum_{lm}\int_{-\infty-i\epsilon}^{\infty-i\epsilon}  
  \!\!\!\! {d\omega}
 \frac{\hat R^{\rm up}_{\omega lm}(r) 
({\cal J}\hat R^{\rm in}_{\omega lm}(r'))}
{k_H A^{\rm in}}\cr
& \quad \times e^{-i\omega(t-t')+im(\varphi-\varphi')} 
  S_{\omega lm}(\theta)S_{\omega lm}(\theta').
  \label{Conti_cancel}
\end{align}
The integration path is now combined to be continuous from $-\infty-i\epsilon$ to $\infty-i\epsilon$, and the asymptotic behavior of the integrand {at a large $|\omega|$} is $\propto \exp(-i\omega(t-t')-i\omega(r_*-r'_*))$. Hence, we can close the integration contour in the lower half complex plane. The pole of $1/A^{\rm in}$ is corresponding to the pole of $1/A^{\rm out}$, {\it i.e.}, 
\begin{align}
 A^{\rm in}_{\bar\omega^*_{nlm} lm}=0. 
\end{align}
Notice that 
\begin{align}
R^*_{nlm}(r) R^*_{nlm}(r')
 =\frac{\hat R^{\rm up}_{\bar\omega_{nlm}^*lm}(r)
 \hat R^{\rm up}_{\bar\omega_{nlm}^*lm}(r')}
  {-2i k_H(\bar\omega^*,m) 
 A^{\rm out}_{\bar\omega^* lm} \partial_{\bar\omega^*}
  A^{\rm in}_{\bar\omega^* lm}},
\end{align}
and 
\begin{align}
{\cal J}\hat R^{\rm in}_{\bar\omega^* lm}
 = \frac{\hat R^{\rm up}_{\bar\omega^* lm}}{{\cal J} A^{\rm in}_{\bar\omega^* lm}} 
 =\frac{R^{\rm up}_{\bar\omega^* lm}}{A^{\rm out}_{\bar\omega^* lm}}. 
\end{align}
The integral \eqref{Conti_cancel} completely cancels the contribution from the term that contains the factor $R^*_{nlm}(r) R^*_{nlm}(r')$ in \eqref{Eq:Green1}. 

Then, the remaining part of $G_R(x,x')$ with $t>t'$ and $r>r'$ is 
\begin{align}
G_{R}&(x,x')
\cr=&
i\biggl[
 \sum_{lm} \int_{-\infty}^{\infty}d\omega
\frac{\hat R^{\rm up}_{\omega lm}(r) 
  \hat R^{\rm in}_{\omega lm}(r')}{4\pi k_H A^{\rm out}}\cr
 & \quad \times e^{-i\omega(t-t')+im(\varphi-\varphi')} 
  S_{\omega lm}(\theta)S_{\omega lm}(\theta')\cr
 &+\sum_n \frac{\hat R^{\rm up}_{\bar\omega_{nlm}lm}(r)
 \hat R^{\rm up}_{\bar\omega_{nlm}lm}(r')}
  {2i k_H(\bar\omega_{nlm},m) 
 A^{\rm in}_{\bar\omega_{nlm} lm} \partial_{\bar\omega_{nlm}}
  A^{\rm out}_{\bar\omega_{nlm} lm}}\cr
& \quad \times  e^{-i\bar\omega_{nlm}(t-t')} 
S_{\bar\omega_{nlm} lm}(\theta)
  S_{\bar\omega_{nlm} lm}(\theta')e^{im(\varphi-\varphi')} 
 \biggr].\cr
\end{align}
In a similar manner, the integration contour in the first term can be changed to cancel the second term to arrive at the expression of the first term in Eq.~\eqref{Eq:Green2}. 
Therefore, we complete the proof of the equivalence of the two forms of retarded Green's function. 
This equivalence shows that the complete set of solutions selected based on the normalizability, which leads to the retarded Green's function given in Eq.~\eqref{Eq:Green1}, is fully consistent with the alternative definition of the retarded Green's function given in Eq.\eqref{Eq:Green2}.

\bibliography{reference}

\end{document}